\documentclass[%
 reprint,
superscriptaddress,
 amsmath,amssymb,
 aps, prl,
floatfix
]{revtex4-2}

\usepackage{lipsum}
\usepackage{physics}
\usepackage{graphicx}
\usepackage{dcolumn}
\usepackage{bm}
\usepackage[colorlinks=true,allcolors=blue]{hyperref}

\begin{document}


\title{Photonic systolic array for all-optical matrix-matrix multiplication}%

\author{Jungmin Kim}
\email{jkim2325@wisc.edu}
\affiliation{Department of Electrical and Computer Engineering, University of Wisconsin-Madison, Madison, WI 53706, USA}
\affiliation{These authors contributed equally.}

\author{Qingyi Zhou}
\email{qzhou75@wisc.edu}
\affiliation{Department of Electrical and Computer Engineering, University of Wisconsin-Madison, Madison, WI 53706, USA}
\affiliation{These authors contributed equally.}

\author{Zongfu Yu}
\affiliation{Department of Electrical and Computer Engineering, University of Wisconsin-Madison, Madison, WI 53706, USA}

\date{\today}

\begin{abstract}
Systolic arrays have proven to be highly efficient for parallelized matrix-matrix multiplication (MMM), utilizing synchronized, heartbeat-like data flows across an array of processing elements. While optical structures such as waveguide crossbar arrays and Mach-Zehnder interferometer-based meshes serve as photonic equivalents to the systolic arrays, the disparity between the two input matrices for multiplication---one using optical signals and the other with system-defined parameters---gives rise to a bottleneck in modern machine-learning tasks, such as evaluating attention scores in large language models. Here, we propose a photonic systolic array that performs MMM entirely with optical signals, utilizing homodyne detection at each array cell. Adjoint-based design of compact on-chip freeform optical modules enables precise control of light flow without bulky waveguide coupling schemes. The operation of a $4\times4$ photonic systolic array is numerically verified, achieving a theoretical computation density of $6.2~\mathrm{PMACs}/\mathrm{mm}^2/\mathrm{s}$. This design marks a significant step toward practical photonic computing hardware for modern AI workloads.
\end{abstract}
\maketitle

\begin{figure*}[t!]
    \centering
    \includegraphics[scale=1]{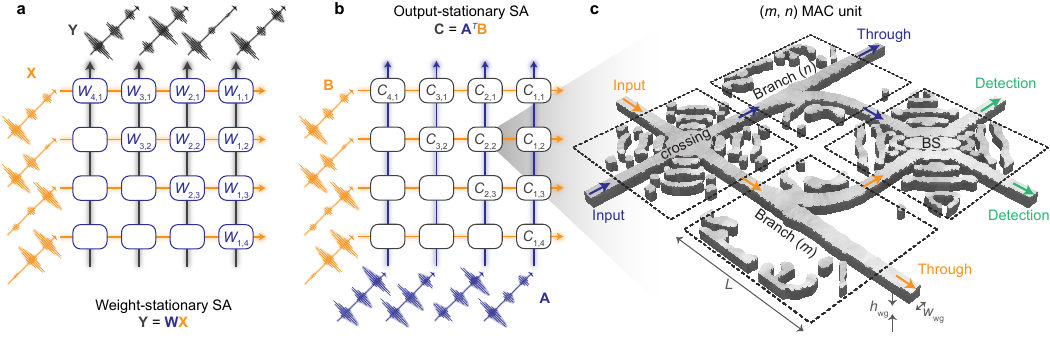}
    \caption{Concept of photonic systolic array. \textbf{a}, Weight-stationary type: input signals $\vb X$ are transformed to the output signals $\vb Y = \vb W \vb X$ by a system parameters $\vb W$. \textbf{b}, Output-stationary type: two input signals ($\vb A$ and $\vb B$) are multiplied on the array of multiply-accumulate (MAC) units, resulting in the stationary output ($\vb C = \vb A^T\vb B$). The array is interconnected via vertical and horizontal waveguides, carrying amplitude-modulated optical pulses for parallel MAC operations, respectively. \textbf{c}, Each optical MAC unit consists of a waveguide crossing, two branches indexed by $m$ and $n$, and a beam splitter for homodyne detection. The waveguides and submodules are built from a Si slab structure with uniform thickness $h_\mathrm{wg}$, embedded within a SiO$_2$ buried oxide layer and cladding. Parameters: $L=3.50$; $h_\text{wg} = 0.22$; $w_\text{wg} = 0.3$ [$\mu\text{m}$]. }
    \label{fig:enter-label}
\end{figure*}

\section*{Introduction}

With the recent rise of large language models \cite{chatgpt2022, Zhao2023, Thirunavukarasu2023, Boiko2023} for generative artificial intelligence, which rely on the attention mechanism involving extensive matrix-matrix multiplication (MMM) operations \cite{Vaswani2017}, there has been an increasing demand for energy-efficient and highly parallelized computing hardware. One notable example is the systolic array (SA), which was originally introduced in 1978 \cite{kung1979} and recently has been revisited for its efficiency in MMM \cite{TPU2017, He2020, Xu2023, Ye2023, Si2024, TPU2024}. As implemented in Google's tensor processing units \cite{TPU2017, TPU2024}, this architecture performs MMM directly on a two-dimensional array of processing elements (PEs), significantly reducing the need for memory access by utilizing the heartbeat-like movement of input and/or output data streams. Importantly, SAs can be classified into two types: weight-stationary (WSSA) and output-stationary (OSSA) \cite{Xu2023}. In WSSA, MMM works similarly to a linear transformation, where a weight matrix ($\vb W$) is preloaded onto the array of PE while an input matrix ($\vb X$) is streamed in one direction, resulting in the output matrix ($\vb Y = \vb W \vb X$) streamed out to another direction, as shown in Fig. 1a. On the other hand, OSSA receives two input matrices ($\vb A$ and $\vb B$) and accumulates the multiplication result ($\vb C = \vb A^T \vb B$) on the spot, as shown in Fig. 1b.

Leveraging photon as an information carrier \cite{McMahon2023}, significant efforts have been devoted to the optical analogy of parallel computing hardwares to address the needs with extreme operation speed and energy efficiency \cite{Lin2018, Hamerly2019, Shastri2021, Zhou2022, Wang2022, Ashtiani2022, Chen2023, Huo2023, Ou2024}. For instance, waveguide crossbar arrays \cite{Feldmann2021, Xu2022, Aggarwal2023, Dong2023, Dong2024, MoralisPegios2024}, unitary meshes based on Mach-Zehnder interferometers (MZI) \cite{Reck1994, Carolan2015, Clements2016, Shen2017, Prez2017, Hughes2018optica}, and inverse-designed nanophotonic structures \cite{Khoram2019, Camacho2021, Nikkhah2024} can be regarded as WSSAs in the sense that only one of two matrices is encoded as traveling optical time signals. That is, the other matrix is not encoded optically, but as near-field (resonant) coupling constants, electro-optic modulation phases at the intersections, or complex material distribution. However, this weight-stationary approach is not ideal for certain machine-learning tasks; during training of deep neural networks, weights need to be not only programmable but also directly mappable to the corresponding site, unlike MZI meshes which requires prior nulling process \cite{Clements2016}. Furthermore, two multiplicands often need to be tuned dynamically, such as the multiplication of query ($\vb Q = \vb W_Q \vb X$) and key ($\vb K = \vb W_K \vb X$) both as a function of embeddings $\vb X$ in transformers \cite{Vaswani2017}. In this context, a photonic OSSA is particularly important, allowing for the general MMM between two optical signals on an equal footing.

Despite its potential advantages, however, photonic OSSA has been explored far less compared to WSSA. Although a promising OSSA prototype has been suggested in Refs. \cite{Youngblood2023, RahimiKari2024} with a single-element demonstration, scalability remains uncertain due to its large footprint and the associated incoherency issue from phase errors. In this work, we advance this architecture by proposing a small-footprint, freeform-designed, output-stationary, silicon-on-insulator photonic systolic array (PSA) for real-valued MMM between purely optical signals. By employing the adjoint state method \cite{LalauKeraly2013, Hughes2018optica, Hughes2018, Hammond2022, Hammond2023, Wu2023, Nikkhah2024}, we precisely target the amplitude and phase-matching conditions necessary for even power distribution and homodyne detection \cite{Hamerly2019, Chen2023}. This approach resolves the remaining incoherency issues upon device scaling in the preliminary studies \cite{Youngblood2023, RahimiKari2024}, while significantly reducing the device footprint down to ten-micron scale per element by eliminating the need for waveguide coupling schemes. As a result, we achieve an exceptionally high computation density, theoretically reaching $6.2~\mathrm{PMACs}/\mathrm{mm}^2/\mathrm{s}$. As an example, we demonstrate the temporal operation of $4\times 4$ PSA, enabled by GPU-accelerated finite-difference time-domain (FDTD) simulation \cite{Tidy3d, Minkov2024}.

\section*{Results}
\subsection*{Operation principle}
Scalar multiplication and addition are the two main arithmatic operations for MMM, which can be performed optically using homodyne detection \cite{Hamerly2019, Chen2023, RahimiKari2024}. This involves a beam splitter (BS) that interferes two monochromatic signals, $ae^{-i\omega_0 t}$ and $ibe^{-i\omega_0 t}$, of a $\pi/2$ phase difference. When these signals impinge into a 50:50 BS, the scattering relation can be expresssed as:
\begin{equation}
    \mqty(o_1 \\ o_2) = \frac{1}{\sqrt 2}\mqty(1 & i \\ i & 1)\mqty(a \\ ib),
\end{equation}
where $a, b$ and $o_{1,2}$ are real-valued input amplitudes and the complex output signals, respectively. The intensity difference between the two output beams is written as $\Delta I\equiv |o_1|^2-|o_2|^2 = -2\Re(a^*b) = -2ab$,
effectively performing scalar multiplication on the two input amplitudes up to a factor of $-2$. Furthermore, if the amplitudes are slowly varying over time as $a(t)$ and $b(t)$, we can approximate the received photon counts difference at two detectors within a time range $\qty[t_s,t_e]$ as
\begin{equation}
    \Delta N = -\frac{2 A}{\hbar\omega_0} \int_{t_s}^{t_e} dt a(t)b(t), 
\end{equation}
where $A$ is the area of detectors. These arriving photons are converted to electrons with quantum efficiency $\eta$ and accumulated at a capacitor, resulting in the charge $\ip{a}{b}_t \equiv \eta q \Delta N_{12}$ as the real-valued inner product between $a$ and $b$ along time. It is noted that throughout the paper, units of intensity (continuous waves) and power (or, energy for amplitude-modulated pulses) can be used interchangeably for the device output.

Now, our approach focuses on (1) designing a compact photonic MAC unit and (2) implementing it in an array to combine temporal inner products with spatial outer products\cite{RahimiKari2024}, to achieve MMM (Fig. 1b). In this scheme, every single column vector of input matrices $(\vb A$ and $\vb B$) is encoded as a time signal $A_m(t)$ and $B_n(t)$ consisting of several optical pulses, where $m$ and $n$ are array indices. Each element of the output matrix is computed through the photonic MAC process.

The design of the MAC unit, illustrated in Fig. 1c, includes four key components: a waveguide crossing, two indexed ($m$ and $n$) branches, and a BS, all of which are free-form designed within the black square area. In each $(m,n)$ MAC unit, the input waves from neighboring cells $(m+1,n)$ and $(m,n+1)$ first cross without crosstalk. The waves then fork into a straight path ($\sim 1-k^{-1}$) and a right-angled path ($\sim k^{-1}$) with the noted split ratio, where $k=m, n$. The branched signals lastly encounter at the BS to return the inner product of the two time signals. Notably, this configuration allows the input signals to travel through waveguides distributing almost the same portion of their energy for detection at each MAC unit.

\begin{figure*}[t!]
    \centering
    \includegraphics[scale=1]{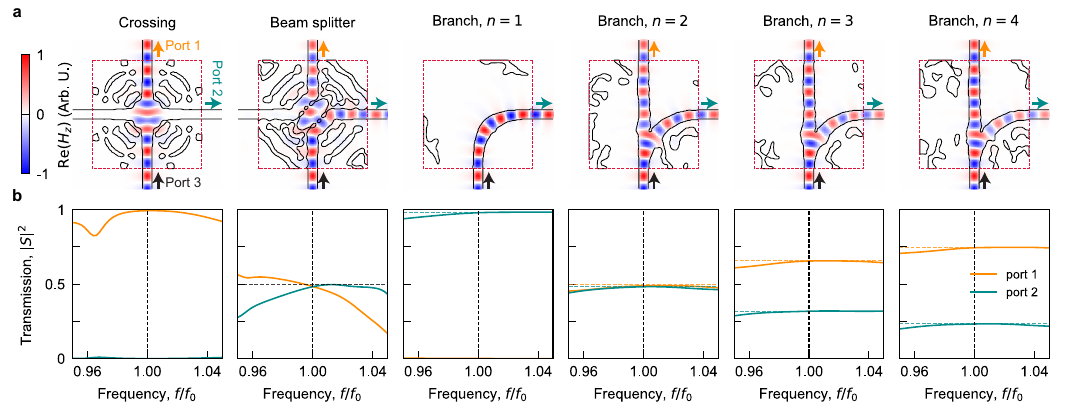}
    \caption{Submodule designs. \textbf{a}, Inverse design results from adjoint optimization: crossing, beam splitter, and branches for $n=1, 2, 3$ and $4$ (from left to right). Black contours and red dashed boxes represent the Si/SiO$_2$ boundary and the square design region with a side length of 3.5 $\mu$m, respectively. The colormap illustrates the wave flow, $\Re(H_z)$, with TM$_{00}$ mode excitation at the bottom port (black arrows, $f=f_0$). \textbf{b}, Transmission spectra of the structures shown in \textbf{a} for two output ports (yellow and teal arrows). Horizontal lines mark the target transmission values at the carrier frequency $f_0 =$ 193.4 THz.}
    \label{fig:enter-label}
\end{figure*}

\subsection*{Adjoint-based submodule design}

As mentioned, the primary goals of the inverse design in this scheme are compactness, equal power distribution, and precise phase matching for interference-based operation. In sharp contrast to the conventional approach relying on waveguide near-field coupling, we utilize adjoint method for the inverse design, by which we can directly assign the necessary specifications for the submodules (e.g., amplitude and phase of the $S$-parameter) while significantly reducing the design footprint.

We assume the TM$_{00}$ mode at the frequency $f_0 = 193.4$ THz of the rectangular waveguide as a carrier mode, which is injected from the bottom port of each submodule (black arrows, Fig. 2b). Then, $S$-parameters for the same mode are measured at the top ($S_{13}$, yellow arrows) and right-side ($S_{23}$, teal arrows) ports for comparison with target values. For the target phase of the $S$-parameters, we set all output phases to converge to $k_\mathrm{eff} L_\mathrm{port}$, where $k_\mathrm{eff}$ and $L_\mathrm{port}$ are the effective wave number through waveguides and the physical port distance, respectively, with the only exception for a $\pi/2$ phase shift at the right-side port of the BS. See Supplementary Fig. S1 for the effective index and mode profile of the waveguide structure.

For the target amplitude of the $S$-parameters, ideal values would be $|S_{23}|^2 = 0$ for the crossing, 0.5 for the BS, and $n^{-1}$ for the $n$-th branch to ensure uniform energy distribution across the array, with $|S_{13}|^2 = 1 - |S_{23}|^2$. However, due to practical limitations on achieving zero insertion loss for broadband operation, known as the Bode-Fano limit \cite{Fano1950}, a bit of margin for the insertion loss is incorporated into the optimizations as $\alpha \equiv |S_{13}|^2 + |S_{23}|^2 \lesssim 1$, in a loss-compensated fan-out design approach \cite{Youngblood2023}. Based on this approach, the practical target amplitudes for branch submodules are slightly adjusted from their ideal values, as depicted by dashed horizontal lines in Fig. 2b. See Supplementary Note S1 for the details of loss compensation.

By specifying the target amplitude and phase for the $S$-parameters, the gradient-descent optimization for each submodule is available, as shown in Fig. 2 (see Supplementary Note S2 and Figs. S2 and S3 for detailed simulation descriptions). The submodule structures and the corresponding wave profiles are shown in Fig. 2a. Importantly, the output phase at ports 1 and 2 (yellow and teal arrows) are all matched, except for the $\pi/2$ phase shift at port 2 of the BS. Figure 2b illustrates the frequency response of the designed structures, showing good agreement between the target transmission levels (dashed lines) and the actual transmission (solid lines). Notably, these transmission spectra provide a rough estimation for operation bandwidth around $0.01f_0$, which will be analyzed in the next section.

\begin{figure*}[t!]
    \centering
    \includegraphics[scale=1]{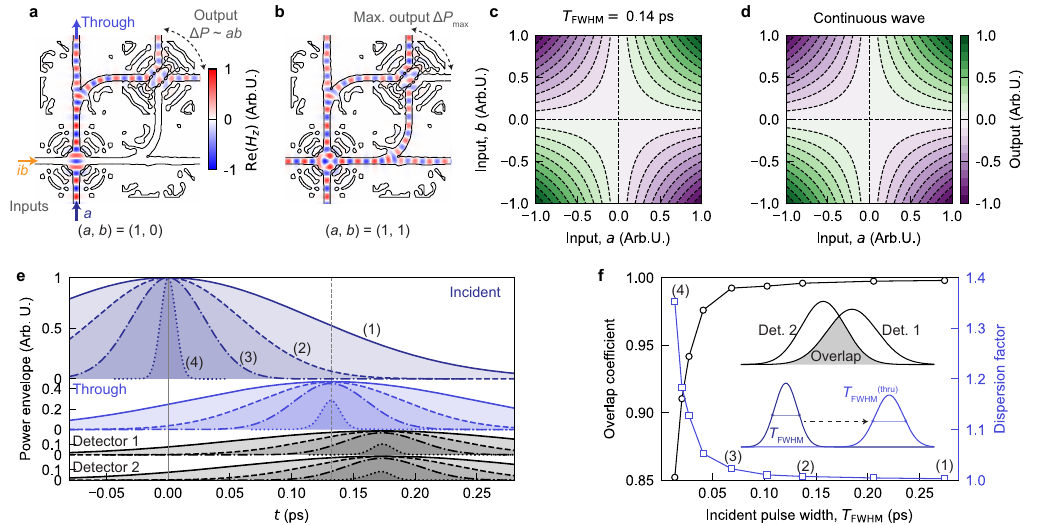}
    \caption{Time-domain operation of the $(2,2)$-MAC unit. \textbf{a},\textbf{b}, Illustration of the $(2,2)$-MAC unit and the corresponding wave propagation for input combinations $(a,b)=(1,0)$ (\textbf{a}) and $(1,1)$ (\textbf{b}) at frequency $f=f_0$. \textbf{c},\textbf{d}, Scalar multiplication results for $-1<a, b<1$ using a finite-width pulse (\textbf{c}) and continuous wave ($f=f_0$, \textbf{d}). \textbf{e}, Evolution of Gaussian pulses from the input (top) to the through (middle) and two detection ports (bottom), with various incident pulse widths (1-4). \textbf{f}, Measure of signal deformation as a function of the incident pulse width: overlap coefficient between two detection signals (black line) and the dispersion factor through the MAC unit (blue line), as illustrated by inset diagrams.}
    \label{fig:enter-label}
\end{figure*}

\subsection*{Time-domain analysis}
The key demonstration of this work is the operation of PSA, which relies on the time-domain streaming of data through optical pulses along waveguides and their exact synchronization. To assess device performance, it is essential to evaluate how different pulse widths affect the operation, thereby determining the maximum achievable bandwidth. We therefore test the $(2,2)$-MAC unit as an example, consisting of one crossing, two branches with $n=m=2$, and one BS, as shown in Fig. 3a (see detailed simulation setup in Supplementary Fig. S4). In this configuration, approximately half of the input power is expected to be passed to subsequent units, while the other half is detected as:
\begin{equation}
    \mqty(o_1 \\ o_2) \approx \frac{1}{\sqrt{2}}\mqty(1 & i \\ i & 1) \frac{1}{\sqrt{2}} \mqty(ib \\ a),
\end{equation}
similar to Eq. (1) except for the transposed inputs after passing through the branches and the factor of $1/\sqrt{2}$, leading to the power difference $\Delta P = |o_1|^2 - |o_2|^2 \approx ab$. We define the maximum power as $P_\mathrm{max} = \Delta P(a=1, b=1)$, as shown in Fig. 3b, which is used for the normalization of the detected power to guarantee that the device output for the input $(1,1)$ corresponds to one.

In Figs. 3c and 3d, we verify that the (2,2)-MAC unit performs scalar multiplication accurately for both finite-width Gaussian pulses $(a,ib)\exp\qty[-(t/T)^2/2]$ with full width at half maximum (FWHM) $T_\mathrm{FWHM}\sim 0.14$ ps (Fig. 3c) and continuous waves for $T\rightarrow \infty$ (Fig. 3d). The colormap displays the normalized device output $\Delta P / P_\mathrm{max}$, which matches the ground truth ($ab$) represented by dashed contours.

However, due to the non-flat spectral responses shown in Fig. 2c, reducing the pulse width $T$ leads to performance degradation. To quantify this, we use Gaussian input pulses with different FWHMs from 0.274 ps to 0.014 ps, corresponding to pulse widths in frequency domain from $f_\mathrm{width} \equiv (2\pi T)^{-1} = f_0/200$ to $f_0/10$, respectively. Fig. 3e shows that as the input pulse (top panel) narrows in time [$T_\mathrm{FWHM} = 0.274$ ps (1) to 0.014 ps (4)], the output pulses at the through port (center panel) and detection ports (bottom panel) become increasingly dispersed, modifying both their heights and shapes. For instance, a dispersion factor, defined as the ratio of the FWHM at the through port to the incident FWHM (i.e., $T_\mathrm{FWHM}^\mathrm{(thru)}/T_\mathrm{FWHM}$), increases rapidly when the incident FWHM drops below 0.14 ps, as shown by the blue curve in Fig. 3f. This dispersion primarily arises from the crossing and branch submodules. Additionally, the BS’s dispersion, illustrated in Fig. 2c, contributes to the discrepancy between the two output pulses at different detection ports. To evaluate the synchronization between the two detected powers, $p_{1,2}(t) = |o_{1,2}(t)|^2$, for the input $(a,b)=(1,0)$, we define the overlap as:
\begin{equation}
    \mathrm{Overlap}(p_1,p_2)\equiv \frac{\int \min\qty[p_1(t), p_2(t)]dt}{\qty[\int p_1(t)dt\int p_2(t)dt]^{1/2}}, 
\end{equation}
As shown in Fig. 3f, the overlap between the two output ports rapidly decreases below one as the pulse width narrows, indicating that the device cannot ensure precise homodyne detection due to the unsynchronized signals. Based on these studies, it is safe to conclude that sufficiently long optical pulses can robustly perform scalar multiplication and inner product without suffering from signal degradation throughout the system.

\begin{figure*}[t!]
    \centering
    \includegraphics[scale=1]{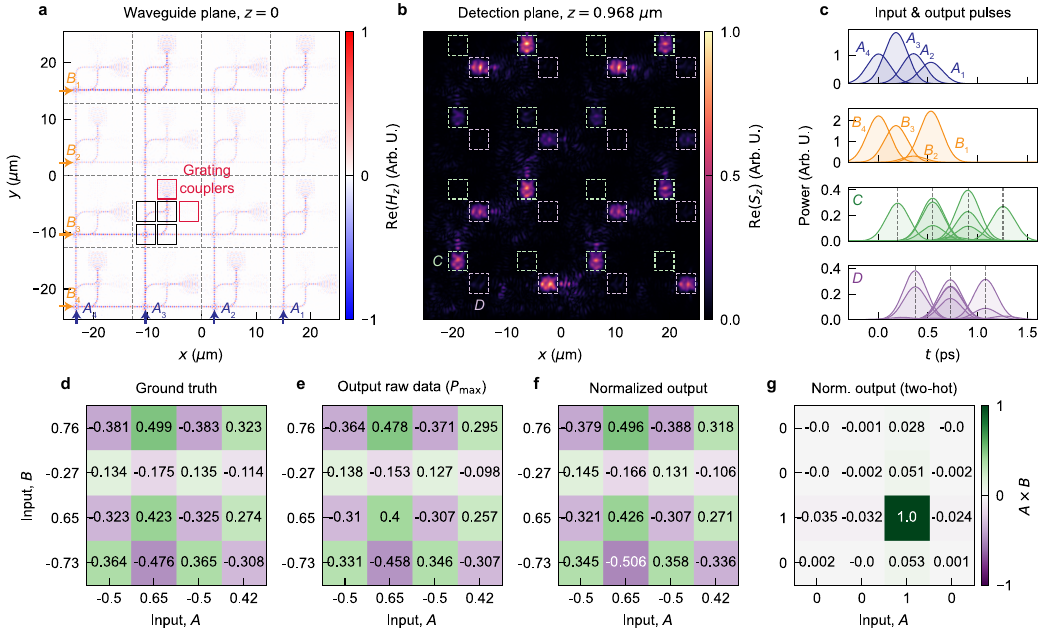}
    \caption{Full-wave simulation of an optical systolic array for the outer product of two vectors. \textbf{a}, Example input (A and B) and output (C and D) signals for a $4\times4$ PSA. Gray dashed lines denote the peak-power time for MAC units where $m+n=8, 7, \cdots, 2$ from left to right. \textbf{b}, Field profile at the waveguide plane. \textbf{c}, Out-of-plane power emission through grating couplers at the detection plane ($z=0.968~\mu\mathrm{m}$, \textbf{c}). Black and red boxes represent the four submodules of the MAC unit and additional grating couplers, respectively. Input ports and output detectors are marked with blue (A) and yellow (B) arrows and green (C) and purple (D) dashed boxes. The lattice period of the array is 12.7 $\mu$m. \textbf{d}-\textbf{g}, Outer product results: ground truth values for given inputs in \textbf{a} (\textbf{d}), output raw data in units of a power constant $P_\mathrm{max}$ defined in the main text (\textbf{e}), and element-wise normalized output for error correction (\textbf{f}); and the element-wise normalized output for inputs $\vb A=[0,1,0,0]$ and $\vb B=[0,0,1,0]$ (\textbf{g}).} 
    \label{fig:enter-label}
\end{figure*}

\subsection*{Spatial extension}
We now extend the concept toward an array of the MAC units based on the previous time-domain analysis, enabling spatial parallelization of a MMM as shown in Fig. 1b. In sharp contrast to the single MAC operation, however, the way of arranging practical detectors within each of elements becomes a problem, since they cannot be simply regarded as in-plane ports in practical setting. Among several options such as integrated on-chip detectors using InGaN/GaN nanowires \cite{Tchernycheva2014} or graphene sheets \cite{Gan2013}, we utilize grating couplers \cite{Hammond2022} to extract optical energy out to $z$-direction in free space, so that we can acquire planar 2D images over the chip, as proposed in Ref. \cite{RahimiKari2024}. Thus, we additionaly inverse design a grating coupler in a similar manner with submodules, obtaining a freeform layout of the same moudle size ($L$) that efficiently converts the waveguide mode into a Gaussian beam of beam waist $L/2$ along $z$-direction. Detailed schematic and the training result is shown in Supplementary Figs. S5 and S6.

Figure 4 demonstrates the outer product $\vb A \otimes \vb B$ (i.e., a MMM only with a single temporal element) using $4\times4$ array of MAC units. In detail, two grating couplers are attached to the two output ports of the BS at each element, as marked by red boxes in Fig. 4a. Then, a single pulse of amplitude $A_m$ and $iB_n$ is injected through each waveguide from bottom and left side, respectively. Those optical signals are processed at the array of MAC units as shown in Fig. 4a, and the output signals are emitted through the grating couplers, which are then recorded by CMOS sensor placed over the chip as illustrated in Fig. 4b. Green and purple dashed boxes indicate the location of the pair of detectors for each MAC unit, having the strong vertical photon flux through them. See Supplementary Fig. S7 for the simulation setup. 

While Figs. 4a and 4b only visualize the $f=f_0$ component of the device operation, indeed the optical pulses with finite pulse widths should be injected in such a way that $A_m$ and $B_m$ encounter right at the $(m,n)$-MAC unit, maximizing the computation accuracy and efficiency. Figure 4c shows the time delay of pulses to ensure this operation. $A_4$ and $B_4$ are initially injected simultaneously, which are followed by $m,n = 3$, 2, and 1, with certain delays corresponding to the traveling time for an optical pulse between two adjacent units. Those input powers are equally distributed over the array and processed, and then output pulses are emitted from the grating couplers turn by turn from the lower left unit ($m+n = 8$) to upper right one ($m+n = 2$).

Figures 4d-4g display the outer product results. For the same input signals $\vb A = [0.42, -0.5, 0.65, -0.5]$ and $\vb B = [0.76, -0.27, 0.65, -0.73]$ as optical pulses with the FWHM of 0.274 ps as shown in Fig. 4c, we obtain the received energy $P_{m,n}^\mathrm{C,D}$) at detectors C and D per each MAC unit, respectively, and calculate the raw output by $\Delta P_{m,n}(\vb A, \vb B) = P_{m,n}^\mathrm{C} - P_{m,n}^\mathrm{D}$ as a function of input $\vb A$ and $\vb B$. The ground truth result and the raw output are compared by Figs. 4d and 4e, respectively, where the raw output is represented in units of normalization constant: $P_\mathrm{max}\equiv\max_{m,n} \Delta P_{m,n}(\vb e_m, \vb e_n)$, where $\vb e_i$ is a one-hot vector with index $i$. In this way, we define the multiplication output ``1" as a maximum physical energy (power) difference achievable over the array with corresponding pulse injections.

While this simple normalization already shows a quite good accuracy, quantified by the mean average error (MAE) of 0.017 with respect to the ground truth table, the element-wise differential normalization by $\Delta P_{m,n}(\vb e_m, \vb e_n)$ provides a more accurate output as displayed in Fig. 4f, with reduced MAE = 0.0097 (see Supplementary Fig. S8 for the normalization powers). This can be realized by a tailored filter between the waveguide and the detection planes or the slight adjustment of the detection time windows. Lastly, the multiplication result $\Delta P(\vb e_2, \vb e_3)$ normalized by $\Delta P_{m,n}(\vb e_m, \vb e_n)$ is shown in Fig. 4g, showcasing the fair errors in relation to the ``multiplication by 0," despite the spread of the error signals along the waveguides due to the reflections from submodules. In addition, we also evaluate the average error of the device with a set of $10^4$ input samples, resulting in the reduction of MAE from 0.0212 to 0.0185, for simple and differential normalization techniques as described in the above (see Supplementary Fig. S9 for details).

\section*{Discussion}
We have only demonstrate the single-pulse operation of a single MAC unit and its extension to $4\times 4$ array for scalar multiplication and the outer product, respectively. However, our system is linear time-invariant without time-varying components, in terms of the relationship between input and output fields. This property enables a straightforward generalization to multiple pulses for the computation of inner products between time signals and matrix-matrix multiplications between two spatio-temporally encoded signals. Consequently, we achieved a single MAC operation within an area of $(12.7~\mu\mathrm{m})^2$ at a data rate of $f_\mathrm{data}\sim 1~\mathrm{THz}$, given that pulse trains can be streamed every 1 ps without overlap, as depicted in Fig. 4c. This corresponds to a theoretical computing density of 6.2 PMACs/mm$^2$/s. 

It is noted that the THz-scale rate is significantly limited by the current operation speed of imaging devices, typically constrained to the MHz frame rate. However, the device remains suitable for processing very long pulse trains. Accumulating photons over an extended period not only reduces errors from photon shot noise and device-level imperfections (Fig. 4f) but also compensates for the lower speed limitations due to the imaging device. This implies the particular advantage of the device in data processing involving very large arrays or matrices.

In summary, we employed the adjoint method to inverse-design the submodules (waveguide crossings, branches, and beam splitters) in a free-form shape, arranging them to constitute the photonic systolic array. The adjoint method allowed for precise targeting of both the amplitudes and phases of the scattering parameters while minimizing the device footprint, which are the key factors for ensuring accurate operation and high computational density. By using finite-difference time-domain simulations, we demonstrated single-unit operation in the time domain and its extension to spatial arrays, achieving matrix-matrix multiplication. Our design enables multiplication between two purely optical signals, in contrast to many existing photonic matrix-matrix or matrix-vector multipliers that operate between an optical signal and electronic modulation such as MZI modulator. Consequently, our approach will significantly advance the energy-efficient machine learning acceleration, benefiting not only inference but also training, especially when dealing with models involving extremely large data sizes.

\begin{acknowledgments}
This work was supported by the Army Research Office through a Multidisciplinary University Research Initiative program (Grant No. W911NF-22-2-0111).
\end{acknowledgments}

\appendix

\bibliography{references}

\begin{thebibliography}{49}%
\makeatletter
\providecommand \@ifxundefined [1]{%
 \@ifx{#1\undefined}
}%
\providecommand \@ifnum [1]{%
 \ifnum #1\expandafter \@firstoftwo
 \else \expandafter \@secondoftwo
 \fi
}%
\providecommand \@ifx [1]{%
 \ifx #1\expandafter \@firstoftwo
 \else \expandafter \@secondoftwo
 \fi
}%
\providecommand \natexlab [1]{#1}%
\providecommand \enquote  [1]{``#1''}%
\providecommand \bibnamefont  [1]{#1}%
\providecommand \bibfnamefont [1]{#1}%
\providecommand \citenamefont [1]{#1}%
\providecommand \href@noop [0]{\@secondoftwo}%
\providecommand \href [0]{\begingroup \@sanitize@url \@href}%
\providecommand \@href[1]{\@@startlink{#1}\@@href}%
\providecommand \@@href[1]{\endgroup#1\@@endlink}%
\providecommand \@sanitize@url [0]{\catcode `\\12\catcode `\$12\catcode `\&12\catcode `\#12\catcode `\^12\catcode `\_12\catcode `\%12\relax}%
\providecommand \@@startlink[1]{}%
\providecommand \@@endlink[0]{}%
\providecommand \url  [0]{\begingroup\@sanitize@url \@url }%
\providecommand \@url [1]{\endgroup\@href {#1}{\urlprefix }}%
\providecommand \urlprefix  [0]{URL }%
\providecommand \Eprint [0]{\href }%
\providecommand \doibase [0]{https://doi.org/}%
\providecommand \selectlanguage [0]{\@gobble}%
\providecommand \bibinfo  [0]{\@secondoftwo}%
\providecommand \bibfield  [0]{\@secondoftwo}%
\providecommand \translation [1]{[#1]}%
\providecommand \BibitemOpen [0]{}%
\providecommand \bibitemStop [0]{}%
\providecommand \bibitemNoStop [0]{.\EOS\space}%
\providecommand \EOS [0]{\spacefactor3000\relax}%
\providecommand \BibitemShut  [1]{\csname bibitem#1\endcsname}%
\let\auto@bib@innerbib\@empty
\bibitem [{\citenamefont {OpenAI}(2022)}]{chatgpt2022}%
  \BibitemOpen
  \bibfield  {author} {\bibinfo {author} {\bibnamefont {OpenAI}},\ }\href@noop {} {\bibinfo {title} {Introducing {ChatGPT}}},\ \bibinfo {howpublished} {\url{https://openai.com/index/chatgpt/}} (\bibinfo {year} {2022}),\ \bibinfo {note} {accessed: \today}\BibitemShut {NoStop}%
\bibitem [{\citenamefont {Zhao}\ \emph {et~al.}(2024)\citenamefont {Zhao}, \citenamefont {Zhou}, \citenamefont {Li}, \citenamefont {Tang}, \citenamefont {Wang}, \citenamefont {Hou}, \citenamefont {Min}, \citenamefont {Zhang}, \citenamefont {Zhang}, \citenamefont {Dong}, \citenamefont {Du}, \citenamefont {Yang}, \citenamefont {Chen}, \citenamefont {Chen}, \citenamefont {Jiang}, \citenamefont {Ren}, \citenamefont {Li}, \citenamefont {Tang}, \citenamefont {Liu}, \citenamefont {Liu}, \citenamefont {Nie},\ and\ \citenamefont {Wen}}]{Zhao2023}%
  \BibitemOpen
  \bibfield  {author} {\bibinfo {author} {\bibfnamefont {W.~X.}\ \bibnamefont {Zhao}}, \bibinfo {author} {\bibfnamefont {K.}~\bibnamefont {Zhou}}, \bibinfo {author} {\bibfnamefont {J.}~\bibnamefont {Li}}, \bibinfo {author} {\bibfnamefont {T.}~\bibnamefont {Tang}}, \bibinfo {author} {\bibfnamefont {X.}~\bibnamefont {Wang}}, \bibinfo {author} {\bibfnamefont {Y.}~\bibnamefont {Hou}}, \bibinfo {author} {\bibfnamefont {Y.}~\bibnamefont {Min}}, \bibinfo {author} {\bibfnamefont {B.}~\bibnamefont {Zhang}}, \bibinfo {author} {\bibfnamefont {J.}~\bibnamefont {Zhang}}, \bibinfo {author} {\bibfnamefont {Z.}~\bibnamefont {Dong}}, \bibinfo {author} {\bibfnamefont {Y.}~\bibnamefont {Du}}, \bibinfo {author} {\bibfnamefont {C.}~\bibnamefont {Yang}}, \bibinfo {author} {\bibfnamefont {Y.}~\bibnamefont {Chen}}, \bibinfo {author} {\bibfnamefont {Z.}~\bibnamefont {Chen}}, \bibinfo {author} {\bibfnamefont {J.}~\bibnamefont {Jiang}}, \bibinfo {author} {\bibfnamefont {R.}~\bibnamefont {Ren}}, \bibinfo {author} {\bibfnamefont
  {Y.}~\bibnamefont {Li}}, \bibinfo {author} {\bibfnamefont {X.}~\bibnamefont {Tang}}, \bibinfo {author} {\bibfnamefont {Z.}~\bibnamefont {Liu}}, \bibinfo {author} {\bibfnamefont {P.}~\bibnamefont {Liu}}, \bibinfo {author} {\bibfnamefont {J.-Y.}\ \bibnamefont {Nie}},\ and\ \bibinfo {author} {\bibfnamefont {J.-R.}\ \bibnamefont {Wen}},\ }\href {https://arxiv.org/abs/2303.18223} {\bibinfo {title} {A survey of large language models}} (\bibinfo {year} {2024}),\ \Eprint {https://arxiv.org/abs/2303.18223} {arXiv:2303.18223 [cs.CL]} \BibitemShut {NoStop}%
\bibitem [{\citenamefont {Thirunavukarasu}\ \emph {et~al.}(2023)\citenamefont {Thirunavukarasu}, \citenamefont {Ting}, \citenamefont {Elangovan}, \citenamefont {Gutierrez}, \citenamefont {Tan},\ and\ \citenamefont {Ting}}]{Thirunavukarasu2023}%
  \BibitemOpen
  \bibfield  {author} {\bibinfo {author} {\bibfnamefont {A.~J.}\ \bibnamefont {Thirunavukarasu}}, \bibinfo {author} {\bibfnamefont {D.~S.~J.}\ \bibnamefont {Ting}}, \bibinfo {author} {\bibfnamefont {K.}~\bibnamefont {Elangovan}}, \bibinfo {author} {\bibfnamefont {L.}~\bibnamefont {Gutierrez}}, \bibinfo {author} {\bibfnamefont {T.~F.}\ \bibnamefont {Tan}},\ and\ \bibinfo {author} {\bibfnamefont {D.~S.~W.}\ \bibnamefont {Ting}},\ }\bibfield  {title} {\bibinfo {title} {Large language models in medicine},\ }\href {https://doi.org/10.1038/s41591-023-02448-8} {\bibfield  {journal} {\bibinfo  {journal} {Nature Medicine}\ }\textbf {\bibinfo {volume} {29}},\ \bibinfo {pages} {1930–1940} (\bibinfo {year} {2023})}\BibitemShut {NoStop}%
\bibitem [{\citenamefont {Boiko}\ \emph {et~al.}(2023)\citenamefont {Boiko}, \citenamefont {MacKnight}, \citenamefont {Kline},\ and\ \citenamefont {Gomes}}]{Boiko2023}%
  \BibitemOpen
  \bibfield  {author} {\bibinfo {author} {\bibfnamefont {D.~A.}\ \bibnamefont {Boiko}}, \bibinfo {author} {\bibfnamefont {R.}~\bibnamefont {MacKnight}}, \bibinfo {author} {\bibfnamefont {B.}~\bibnamefont {Kline}},\ and\ \bibinfo {author} {\bibfnamefont {G.}~\bibnamefont {Gomes}},\ }\bibfield  {title} {\bibinfo {title} {Autonomous chemical research with large language models},\ }\href {https://doi.org/10.1038/s41586-023-06792-0} {\bibfield  {journal} {\bibinfo  {journal} {Nature}\ }\textbf {\bibinfo {volume} {624}},\ \bibinfo {pages} {570–578} (\bibinfo {year} {2023})}\BibitemShut {NoStop}%
\bibitem [{\citenamefont {Vaswani}\ \emph {et~al.}(2017)\citenamefont {Vaswani}, \citenamefont {Shazeer}, \citenamefont {Parmar}, \citenamefont {Uszkoreit}, \citenamefont {Jones}, \citenamefont {Gomez}, \citenamefont {Kaiser},\ and\ \citenamefont {Polosukhin}}]{Vaswani2017}%
  \BibitemOpen
  \bibfield  {author} {\bibinfo {author} {\bibfnamefont {A.}~\bibnamefont {Vaswani}}, \bibinfo {author} {\bibfnamefont {N.}~\bibnamefont {Shazeer}}, \bibinfo {author} {\bibfnamefont {N.}~\bibnamefont {Parmar}}, \bibinfo {author} {\bibfnamefont {J.}~\bibnamefont {Uszkoreit}}, \bibinfo {author} {\bibfnamefont {L.}~\bibnamefont {Jones}}, \bibinfo {author} {\bibfnamefont {A.~N.}\ \bibnamefont {Gomez}}, \bibinfo {author} {\bibfnamefont {L.~u.}\ \bibnamefont {Kaiser}},\ and\ \bibinfo {author} {\bibfnamefont {I.}~\bibnamefont {Polosukhin}},\ }\bibfield  {title} {\bibinfo {title} {Attention is all you need},\ }in\ \href {https://proceedings.neurips.cc/paper_files/paper/2017/file/3f5ee243547dee91fbd053c1c4a845aa-Paper.pdf} {\emph {\bibinfo {booktitle} {Advances in Neural Information Processing Systems}}},\ Vol.~\bibinfo {volume} {30},\ \bibinfo {editor} {edited by\ \bibinfo {editor} {\bibfnamefont {I.}~\bibnamefont {Guyon}}, \bibinfo {editor} {\bibfnamefont {U.~V.}\ \bibnamefont {Luxburg}}, \bibinfo {editor}
  {\bibfnamefont {S.}~\bibnamefont {Bengio}}, \bibinfo {editor} {\bibfnamefont {H.}~\bibnamefont {Wallach}}, \bibinfo {editor} {\bibfnamefont {R.}~\bibnamefont {Fergus}}, \bibinfo {editor} {\bibfnamefont {S.}~\bibnamefont {Vishwanathan}},\ and\ \bibinfo {editor} {\bibfnamefont {R.}~\bibnamefont {Garnett}}}\ (\bibinfo  {publisher} {Curran Associates, Inc.},\ \bibinfo {year} {2017})\BibitemShut {NoStop}%
\bibitem [{\citenamefont {Kung}\ and\ \citenamefont {Leiserson}(1979)}]{kung1979}%
  \BibitemOpen
  \bibfield  {author} {\bibinfo {author} {\bibfnamefont {H.~T.}\ \bibnamefont {Kung}}\ and\ \bibinfo {author} {\bibfnamefont {C.~E.}\ \bibnamefont {Leiserson}},\ }\bibfield  {title} {\bibinfo {title} {Systolic arrays (for {VLSI})},\ }in\ \href@noop {} {\emph {\bibinfo {booktitle} {Sparse Matrix Proceedings 1978}}},\ Vol.~\bibinfo {volume} {1}\ (\bibinfo {organization} {Society for industrial and applied mathematics Philadelphia, PA, USA},\ \bibinfo {year} {1979})\ pp.\ \bibinfo {pages} {256--282}\BibitemShut {NoStop}%
\bibitem [{\citenamefont {Jouppi}\ \emph {et~al.}(2017)\citenamefont {Jouppi}, \citenamefont {Young}, \citenamefont {Patil}, \citenamefont {Patterson}, \citenamefont {Agrawal}, \citenamefont {Bajwa}, \citenamefont {Bates}, \citenamefont {Bhatia}, \citenamefont {Boden}, \citenamefont {Borchers}, \citenamefont {Boyle}, \citenamefont {luc Cantin}, \citenamefont {Chao}, \citenamefont {Clark}, \citenamefont {Coriell}, \citenamefont {Daley}, \citenamefont {Dau}, \citenamefont {Dean}, \citenamefont {Gelb}, \citenamefont {Ghaemmaghami}, \citenamefont {Gottipati}, \citenamefont {Gulland}, \citenamefont {Hagmann}, \citenamefont {Ho}, \citenamefont {Hogberg}, \citenamefont {Hu}, \citenamefont {Hundt}, \citenamefont {Hurt}, \citenamefont {Ibarz}, \citenamefont {Jaffey}, \citenamefont {Jaworski}, \citenamefont {Kaplan}, \citenamefont {Khaitan}, \citenamefont {Koch}, \citenamefont {Kumar}, \citenamefont {Lacy}, \citenamefont {Laudon}, \citenamefont {Law}, \citenamefont {Le}, \citenamefont {Leary}, \citenamefont {Liu},
  \citenamefont {Lucke}, \citenamefont {Lundin}, \citenamefont {MacKean}, \citenamefont {Maggiore}, \citenamefont {Mahony}, \citenamefont {Miller}, \citenamefont {Nagarajan}, \citenamefont {Narayanaswami}, \citenamefont {Ni}, \citenamefont {Nix}, \citenamefont {Norrie}, \citenamefont {Omernick}, \citenamefont {Penukonda}, \citenamefont {Phelps}, \citenamefont {Ross}, \citenamefont {Ross}, \citenamefont {Salek}, \citenamefont {Samadiani}, \citenamefont {Severn}, \citenamefont {Sizikov}, \citenamefont {Snelham}, \citenamefont {Souter}, \citenamefont {Steinberg}, \citenamefont {Swing}, \citenamefont {Tan}, \citenamefont {Thorson}, \citenamefont {Tian}, \citenamefont {Toma}, \citenamefont {Tuttle}, \citenamefont {Vasudevan}, \citenamefont {Walter}, \citenamefont {Wang}, \citenamefont {Wilcox},\ and\ \citenamefont {Yoon}}]{TPU2017}%
  \BibitemOpen
  \bibfield  {author} {\bibinfo {author} {\bibfnamefont {N.~P.}\ \bibnamefont {Jouppi}}, \bibinfo {author} {\bibfnamefont {C.}~\bibnamefont {Young}}, \bibinfo {author} {\bibfnamefont {N.}~\bibnamefont {Patil}}, \bibinfo {author} {\bibfnamefont {D.}~\bibnamefont {Patterson}}, \bibinfo {author} {\bibfnamefont {G.}~\bibnamefont {Agrawal}}, \bibinfo {author} {\bibfnamefont {R.}~\bibnamefont {Bajwa}}, \bibinfo {author} {\bibfnamefont {S.}~\bibnamefont {Bates}}, \bibinfo {author} {\bibfnamefont {S.}~\bibnamefont {Bhatia}}, \bibinfo {author} {\bibfnamefont {N.}~\bibnamefont {Boden}}, \bibinfo {author} {\bibfnamefont {A.}~\bibnamefont {Borchers}}, \bibinfo {author} {\bibfnamefont {R.}~\bibnamefont {Boyle}}, \bibinfo {author} {\bibfnamefont {P.}~\bibnamefont {luc Cantin}}, \bibinfo {author} {\bibfnamefont {C.}~\bibnamefont {Chao}}, \bibinfo {author} {\bibfnamefont {C.}~\bibnamefont {Clark}}, \bibinfo {author} {\bibfnamefont {J.}~\bibnamefont {Coriell}}, \bibinfo {author} {\bibfnamefont {M.}~\bibnamefont {Daley}},
  \bibinfo {author} {\bibfnamefont {M.}~\bibnamefont {Dau}}, \bibinfo {author} {\bibfnamefont {J.}~\bibnamefont {Dean}}, \bibinfo {author} {\bibfnamefont {B.}~\bibnamefont {Gelb}}, \bibinfo {author} {\bibfnamefont {T.~V.}\ \bibnamefont {Ghaemmaghami}}, \bibinfo {author} {\bibfnamefont {R.}~\bibnamefont {Gottipati}}, \bibinfo {author} {\bibfnamefont {W.}~\bibnamefont {Gulland}}, \bibinfo {author} {\bibfnamefont {R.}~\bibnamefont {Hagmann}}, \bibinfo {author} {\bibfnamefont {C.~R.}\ \bibnamefont {Ho}}, \bibinfo {author} {\bibfnamefont {D.}~\bibnamefont {Hogberg}}, \bibinfo {author} {\bibfnamefont {J.}~\bibnamefont {Hu}}, \bibinfo {author} {\bibfnamefont {R.}~\bibnamefont {Hundt}}, \bibinfo {author} {\bibfnamefont {D.}~\bibnamefont {Hurt}}, \bibinfo {author} {\bibfnamefont {J.}~\bibnamefont {Ibarz}}, \bibinfo {author} {\bibfnamefont {A.}~\bibnamefont {Jaffey}}, \bibinfo {author} {\bibfnamefont {A.}~\bibnamefont {Jaworski}}, \bibinfo {author} {\bibfnamefont {A.}~\bibnamefont {Kaplan}}, \bibinfo {author}
  {\bibfnamefont {H.}~\bibnamefont {Khaitan}}, \bibinfo {author} {\bibfnamefont {A.}~\bibnamefont {Koch}}, \bibinfo {author} {\bibfnamefont {N.}~\bibnamefont {Kumar}}, \bibinfo {author} {\bibfnamefont {S.}~\bibnamefont {Lacy}}, \bibinfo {author} {\bibfnamefont {J.}~\bibnamefont {Laudon}}, \bibinfo {author} {\bibfnamefont {J.}~\bibnamefont {Law}}, \bibinfo {author} {\bibfnamefont {D.}~\bibnamefont {Le}}, \bibinfo {author} {\bibfnamefont {C.}~\bibnamefont {Leary}}, \bibinfo {author} {\bibfnamefont {Z.}~\bibnamefont {Liu}}, \bibinfo {author} {\bibfnamefont {K.}~\bibnamefont {Lucke}}, \bibinfo {author} {\bibfnamefont {A.}~\bibnamefont {Lundin}}, \bibinfo {author} {\bibfnamefont {G.}~\bibnamefont {MacKean}}, \bibinfo {author} {\bibfnamefont {A.}~\bibnamefont {Maggiore}}, \bibinfo {author} {\bibfnamefont {M.}~\bibnamefont {Mahony}}, \bibinfo {author} {\bibfnamefont {K.}~\bibnamefont {Miller}}, \bibinfo {author} {\bibfnamefont {R.}~\bibnamefont {Nagarajan}}, \bibinfo {author} {\bibfnamefont {R.}~\bibnamefont
  {Narayanaswami}}, \bibinfo {author} {\bibfnamefont {R.}~\bibnamefont {Ni}}, \bibinfo {author} {\bibfnamefont {K.}~\bibnamefont {Nix}}, \bibinfo {author} {\bibfnamefont {T.}~\bibnamefont {Norrie}}, \bibinfo {author} {\bibfnamefont {M.}~\bibnamefont {Omernick}}, \bibinfo {author} {\bibfnamefont {N.}~\bibnamefont {Penukonda}}, \bibinfo {author} {\bibfnamefont {A.}~\bibnamefont {Phelps}}, \bibinfo {author} {\bibfnamefont {J.}~\bibnamefont {Ross}}, \bibinfo {author} {\bibfnamefont {M.}~\bibnamefont {Ross}}, \bibinfo {author} {\bibfnamefont {A.}~\bibnamefont {Salek}}, \bibinfo {author} {\bibfnamefont {E.}~\bibnamefont {Samadiani}}, \bibinfo {author} {\bibfnamefont {C.}~\bibnamefont {Severn}}, \bibinfo {author} {\bibfnamefont {G.}~\bibnamefont {Sizikov}}, \bibinfo {author} {\bibfnamefont {M.}~\bibnamefont {Snelham}}, \bibinfo {author} {\bibfnamefont {J.}~\bibnamefont {Souter}}, \bibinfo {author} {\bibfnamefont {D.}~\bibnamefont {Steinberg}}, \bibinfo {author} {\bibfnamefont {A.}~\bibnamefont {Swing}}, \bibinfo
  {author} {\bibfnamefont {M.}~\bibnamefont {Tan}}, \bibinfo {author} {\bibfnamefont {G.}~\bibnamefont {Thorson}}, \bibinfo {author} {\bibfnamefont {B.}~\bibnamefont {Tian}}, \bibinfo {author} {\bibfnamefont {H.}~\bibnamefont {Toma}}, \bibinfo {author} {\bibfnamefont {E.}~\bibnamefont {Tuttle}}, \bibinfo {author} {\bibfnamefont {V.}~\bibnamefont {Vasudevan}}, \bibinfo {author} {\bibfnamefont {R.}~\bibnamefont {Walter}}, \bibinfo {author} {\bibfnamefont {W.}~\bibnamefont {Wang}}, \bibinfo {author} {\bibfnamefont {E.}~\bibnamefont {Wilcox}},\ and\ \bibinfo {author} {\bibfnamefont {D.~H.}\ \bibnamefont {Yoon}},\ }\href {https://arxiv.org/abs/1704.04760} {\bibinfo {title} {In-datacenter performance analysis of a tensor processing unit}} (\bibinfo {year} {2017}),\ \Eprint {https://arxiv.org/abs/1704.04760} {arXiv:1704.04760 [cs.AR]} \BibitemShut {NoStop}%
\bibitem [{\citenamefont {He}\ \emph {et~al.}(2020)\citenamefont {He}, \citenamefont {Pal}, \citenamefont {Amarnath}, \citenamefont {Feng}, \citenamefont {Park}, \citenamefont {Rovinski}, \citenamefont {Ye}, \citenamefont {Chen}, \citenamefont {Dreslinski},\ and\ \citenamefont {Mudge}}]{He2020}%
  \BibitemOpen
  \bibfield  {author} {\bibinfo {author} {\bibfnamefont {X.}~\bibnamefont {He}}, \bibinfo {author} {\bibfnamefont {S.}~\bibnamefont {Pal}}, \bibinfo {author} {\bibfnamefont {A.}~\bibnamefont {Amarnath}}, \bibinfo {author} {\bibfnamefont {S.}~\bibnamefont {Feng}}, \bibinfo {author} {\bibfnamefont {D.-H.}\ \bibnamefont {Park}}, \bibinfo {author} {\bibfnamefont {A.}~\bibnamefont {Rovinski}}, \bibinfo {author} {\bibfnamefont {H.}~\bibnamefont {Ye}}, \bibinfo {author} {\bibfnamefont {Y.}~\bibnamefont {Chen}}, \bibinfo {author} {\bibfnamefont {R.}~\bibnamefont {Dreslinski}},\ and\ \bibinfo {author} {\bibfnamefont {T.}~\bibnamefont {Mudge}},\ }\bibfield  {title} {\bibinfo {title} {Sparse-tpu: adapting systolic arrays for sparse matrices},\ }in\ \href {https://doi.org/10.1145/3392717.3392751} {\emph {\bibinfo {booktitle} {Proceedings of the 34th ACM International Conference on Supercomputing}}},\ \bibinfo {series and number} {ICS '20}\ (\bibinfo  {publisher} {Association for Computing Machinery},\ \bibinfo {address}
  {New York, NY, USA},\ \bibinfo {year} {2020})\BibitemShut {NoStop}%
\bibitem [{\citenamefont {Xu}\ \emph {et~al.}(2023)\citenamefont {Xu}, \citenamefont {Ma}, \citenamefont {Guo},\ and\ \citenamefont {Li}}]{Xu2023}%
  \BibitemOpen
  \bibfield  {author} {\bibinfo {author} {\bibfnamefont {R.}~\bibnamefont {Xu}}, \bibinfo {author} {\bibfnamefont {S.}~\bibnamefont {Ma}}, \bibinfo {author} {\bibfnamefont {Y.}~\bibnamefont {Guo}},\ and\ \bibinfo {author} {\bibfnamefont {D.}~\bibnamefont {Li}},\ }\bibfield  {title} {\bibinfo {title} {A survey of design and optimization for systolic array-based dnn accelerators},\ }\bibfield  {journal} {\bibinfo  {journal} {ACM Comput. Surv.}\ }\textbf {\bibinfo {volume} {56}},\ \href {https://doi.org/10.1145/3604802} {10.1145/3604802} (\bibinfo {year} {2023})\BibitemShut {NoStop}%
\bibitem [{\citenamefont {Ye}\ \emph {et~al.}(2023)\citenamefont {Ye}, \citenamefont {Zhou}, \citenamefont {Zhou}, \citenamefont {Chen},\ and\ \citenamefont {Li}}]{Ye2023}%
  \BibitemOpen
  \bibfield  {author} {\bibinfo {author} {\bibfnamefont {W.}~\bibnamefont {Ye}}, \bibinfo {author} {\bibfnamefont {X.}~\bibnamefont {Zhou}}, \bibinfo {author} {\bibfnamefont {J.}~\bibnamefont {Zhou}}, \bibinfo {author} {\bibfnamefont {C.}~\bibnamefont {Chen}},\ and\ \bibinfo {author} {\bibfnamefont {K.}~\bibnamefont {Li}},\ }\bibfield  {title} {\bibinfo {title} {Accelerating attention mechanism on fpgas based on efficient reconfigurable systolic array},\ }\bibfield  {journal} {\bibinfo  {journal} {ACM Trans. Embed. Comput. Syst.}\ }\textbf {\bibinfo {volume} {22}},\ \href {https://doi.org/10.1145/3549937} {10.1145/3549937} (\bibinfo {year} {2023})\BibitemShut {NoStop}%
\bibitem [{\citenamefont {Si}\ \emph {et~al.}(2024)\citenamefont {Si}, \citenamefont {Zhang}, \citenamefont {Zhao}, \citenamefont {Lin}, \citenamefont {Xu}, \citenamefont {Xu}, \citenamefont {Liu}, \citenamefont {Jiang}, \citenamefont {Peng},\ and\ \citenamefont {Zhang}}]{Si2024}%
  \BibitemOpen
  \bibfield  {author} {\bibinfo {author} {\bibfnamefont {J.}~\bibnamefont {Si}}, \bibinfo {author} {\bibfnamefont {P.}~\bibnamefont {Zhang}}, \bibinfo {author} {\bibfnamefont {C.}~\bibnamefont {Zhao}}, \bibinfo {author} {\bibfnamefont {D.}~\bibnamefont {Lin}}, \bibinfo {author} {\bibfnamefont {L.}~\bibnamefont {Xu}}, \bibinfo {author} {\bibfnamefont {H.}~\bibnamefont {Xu}}, \bibinfo {author} {\bibfnamefont {L.}~\bibnamefont {Liu}}, \bibinfo {author} {\bibfnamefont {J.}~\bibnamefont {Jiang}}, \bibinfo {author} {\bibfnamefont {L.-M.}\ \bibnamefont {Peng}},\ and\ \bibinfo {author} {\bibfnamefont {Z.}~\bibnamefont {Zhang}},\ }\bibfield  {title} {\bibinfo {title} {A carbon-nanotube-based tensor processing unit},\ }\href {https://doi.org/10.1038/s41928-024-01211-2} {\bibfield  {journal} {\bibinfo  {journal} {Nature Electronics}\ }\textbf {\bibinfo {volume} {7}},\ \bibinfo {pages} {684–693} (\bibinfo {year} {2024})}\BibitemShut {NoStop}%
\bibitem [{\citenamefont {Vahdat}(2024)}]{TPU2024}%
  \BibitemOpen
  \bibfield  {author} {\bibinfo {author} {\bibfnamefont {A.}~\bibnamefont {Vahdat}},\ }\href@noop {} {\bibinfo {title} {Announcing {Trillium}, the sixth generation of {Google Cloud TPU}}},\ \bibinfo {howpublished} {\url{https://cloud.google.com/blog/products/compute/introducing-trillium-6th-gen-tpus}} (\bibinfo {year} {2024}),\ \bibinfo {note} {accessed: \today}\BibitemShut {NoStop}%
\bibitem [{\citenamefont {McMahon}(2023)}]{McMahon2023}%
  \BibitemOpen
  \bibfield  {author} {\bibinfo {author} {\bibfnamefont {P.~L.}\ \bibnamefont {McMahon}},\ }\bibfield  {title} {\bibinfo {title} {The physics of optical computing},\ }\href {https://doi.org/10.1038/s42254-023-00645-5} {\bibfield  {journal} {\bibinfo  {journal} {Nature Reviews Physics}\ }\textbf {\bibinfo {volume} {5}},\ \bibinfo {pages} {717–734} (\bibinfo {year} {2023})}\BibitemShut {NoStop}%
\bibitem [{\citenamefont {Lin}\ \emph {et~al.}(2018)\citenamefont {Lin}, \citenamefont {Rivenson}, \citenamefont {Yardimci}, \citenamefont {Veli}, \citenamefont {Luo}, \citenamefont {Jarrahi},\ and\ \citenamefont {Ozcan}}]{Lin2018}%
  \BibitemOpen
  \bibfield  {author} {\bibinfo {author} {\bibfnamefont {X.}~\bibnamefont {Lin}}, \bibinfo {author} {\bibfnamefont {Y.}~\bibnamefont {Rivenson}}, \bibinfo {author} {\bibfnamefont {N.~T.}\ \bibnamefont {Yardimci}}, \bibinfo {author} {\bibfnamefont {M.}~\bibnamefont {Veli}}, \bibinfo {author} {\bibfnamefont {Y.}~\bibnamefont {Luo}}, \bibinfo {author} {\bibfnamefont {M.}~\bibnamefont {Jarrahi}},\ and\ \bibinfo {author} {\bibfnamefont {A.}~\bibnamefont {Ozcan}},\ }\bibfield  {title} {\bibinfo {title} {All-optical machine learning using diffractive deep neural networks},\ }\href {https://doi.org/10.1126/science.aat8084} {\bibfield  {journal} {\bibinfo  {journal} {Science}\ }\textbf {\bibinfo {volume} {361}},\ \bibinfo {pages} {1004–1008} (\bibinfo {year} {2018})}\BibitemShut {NoStop}%
\bibitem [{\citenamefont {Hamerly}\ \emph {et~al.}(2019)\citenamefont {Hamerly}, \citenamefont {Bernstein}, \citenamefont {Sludds}, \citenamefont {Solja\ifmmode \check{c}\else \v{c}\fi{}i\ifmmode~\acute{c}\else \'{c}\fi{}},\ and\ \citenamefont {Englund}}]{Hamerly2019}%
  \BibitemOpen
  \bibfield  {author} {\bibinfo {author} {\bibfnamefont {R.}~\bibnamefont {Hamerly}}, \bibinfo {author} {\bibfnamefont {L.}~\bibnamefont {Bernstein}}, \bibinfo {author} {\bibfnamefont {A.}~\bibnamefont {Sludds}}, \bibinfo {author} {\bibfnamefont {M.}~\bibnamefont {Solja\ifmmode \check{c}\else \v{c}\fi{}i\ifmmode~\acute{c}\else \'{c}\fi{}}},\ and\ \bibinfo {author} {\bibfnamefont {D.}~\bibnamefont {Englund}},\ }\bibfield  {title} {\bibinfo {title} {Large-scale optical neural networks based on photoelectric multiplication},\ }\href {https://doi.org/10.1103/PhysRevX.9.021032} {\bibfield  {journal} {\bibinfo  {journal} {Phys. Rev. X}\ }\textbf {\bibinfo {volume} {9}},\ \bibinfo {pages} {021032} (\bibinfo {year} {2019})}\BibitemShut {NoStop}%
\bibitem [{\citenamefont {Shastri}\ \emph {et~al.}(2021)\citenamefont {Shastri}, \citenamefont {Tait}, \citenamefont {Ferreira~de Lima}, \citenamefont {Pernice}, \citenamefont {Bhaskaran}, \citenamefont {Wright},\ and\ \citenamefont {Prucnal}}]{Shastri2021}%
  \BibitemOpen
  \bibfield  {author} {\bibinfo {author} {\bibfnamefont {B.~J.}\ \bibnamefont {Shastri}}, \bibinfo {author} {\bibfnamefont {A.~N.}\ \bibnamefont {Tait}}, \bibinfo {author} {\bibfnamefont {T.}~\bibnamefont {Ferreira~de Lima}}, \bibinfo {author} {\bibfnamefont {W.~H.~P.}\ \bibnamefont {Pernice}}, \bibinfo {author} {\bibfnamefont {H.}~\bibnamefont {Bhaskaran}}, \bibinfo {author} {\bibfnamefont {C.~D.}\ \bibnamefont {Wright}},\ and\ \bibinfo {author} {\bibfnamefont {P.~R.}\ \bibnamefont {Prucnal}},\ }\bibfield  {title} {\bibinfo {title} {Photonics for artificial intelligence and neuromorphic computing},\ }\href {https://doi.org/10.1038/s41566-020-00754-y} {\bibfield  {journal} {\bibinfo  {journal} {Nature Photonics}\ }\textbf {\bibinfo {volume} {15}},\ \bibinfo {pages} {102–114} (\bibinfo {year} {2021})}\BibitemShut {NoStop}%
\bibitem [{\citenamefont {Zhou}\ \emph {et~al.}(2022)\citenamefont {Zhou}, \citenamefont {Dong}, \citenamefont {Cheng}, \citenamefont {Dong}, \citenamefont {Huang}, \citenamefont {Shen}, \citenamefont {Zhang}, \citenamefont {Gu}, \citenamefont {Qian}, \citenamefont {Chen}, \citenamefont {Ruan},\ and\ \citenamefont {Zhang}}]{Zhou2022}%
  \BibitemOpen
  \bibfield  {author} {\bibinfo {author} {\bibfnamefont {H.}~\bibnamefont {Zhou}}, \bibinfo {author} {\bibfnamefont {J.}~\bibnamefont {Dong}}, \bibinfo {author} {\bibfnamefont {J.}~\bibnamefont {Cheng}}, \bibinfo {author} {\bibfnamefont {W.}~\bibnamefont {Dong}}, \bibinfo {author} {\bibfnamefont {C.}~\bibnamefont {Huang}}, \bibinfo {author} {\bibfnamefont {Y.}~\bibnamefont {Shen}}, \bibinfo {author} {\bibfnamefont {Q.}~\bibnamefont {Zhang}}, \bibinfo {author} {\bibfnamefont {M.}~\bibnamefont {Gu}}, \bibinfo {author} {\bibfnamefont {C.}~\bibnamefont {Qian}}, \bibinfo {author} {\bibfnamefont {H.}~\bibnamefont {Chen}}, \bibinfo {author} {\bibfnamefont {Z.}~\bibnamefont {Ruan}},\ and\ \bibinfo {author} {\bibfnamefont {X.}~\bibnamefont {Zhang}},\ }\bibfield  {title} {\bibinfo {title} {Photonic matrix multiplication lights up photonic accelerator and beyond},\ }\bibfield  {journal} {\bibinfo  {journal} {Light: Science \& Applications}\ }\textbf {\bibinfo {volume} {11}},\ \href
  {https://doi.org/10.1038/s41377-022-00717-8} {10.1038/s41377-022-00717-8} (\bibinfo {year} {2022})\BibitemShut {NoStop}%
\bibitem [{\citenamefont {Wang}\ \emph {et~al.}(2022)\citenamefont {Wang}, \citenamefont {Ma}, \citenamefont {Wright}, \citenamefont {Onodera}, \citenamefont {Richard},\ and\ \citenamefont {McMahon}}]{Wang2022}%
  \BibitemOpen
  \bibfield  {author} {\bibinfo {author} {\bibfnamefont {T.}~\bibnamefont {Wang}}, \bibinfo {author} {\bibfnamefont {S.-Y.}\ \bibnamefont {Ma}}, \bibinfo {author} {\bibfnamefont {L.~G.}\ \bibnamefont {Wright}}, \bibinfo {author} {\bibfnamefont {T.}~\bibnamefont {Onodera}}, \bibinfo {author} {\bibfnamefont {B.~C.}\ \bibnamefont {Richard}},\ and\ \bibinfo {author} {\bibfnamefont {P.~L.}\ \bibnamefont {McMahon}},\ }\bibfield  {title} {\bibinfo {title} {An optical neural network using less than 1 photon per multiplication},\ }\bibfield  {journal} {\bibinfo  {journal} {Nature Communications}\ }\textbf {\bibinfo {volume} {13}},\ \href {https://doi.org/10.1038/s41467-021-27774-8} {10.1038/s41467-021-27774-8} (\bibinfo {year} {2022})\BibitemShut {NoStop}%
\bibitem [{\citenamefont {Ashtiani}\ \emph {et~al.}(2022)\citenamefont {Ashtiani}, \citenamefont {Geers},\ and\ \citenamefont {Aflatouni}}]{Ashtiani2022}%
  \BibitemOpen
  \bibfield  {author} {\bibinfo {author} {\bibfnamefont {F.}~\bibnamefont {Ashtiani}}, \bibinfo {author} {\bibfnamefont {A.~J.}\ \bibnamefont {Geers}},\ and\ \bibinfo {author} {\bibfnamefont {F.}~\bibnamefont {Aflatouni}},\ }\bibfield  {title} {\bibinfo {title} {An on-chip photonic deep neural network for image classification},\ }\href {https://doi.org/10.1038/s41586-022-04714-0} {\bibfield  {journal} {\bibinfo  {journal} {Nature}\ }\textbf {\bibinfo {volume} {606}},\ \bibinfo {pages} {501–506} (\bibinfo {year} {2022})}\BibitemShut {NoStop}%
\bibitem [{\citenamefont {Chen}\ \emph {et~al.}(2023)\citenamefont {Chen}, \citenamefont {Sludds}, \citenamefont {Davis}, \citenamefont {Christen}, \citenamefont {Bernstein}, \citenamefont {Ateshian}, \citenamefont {Heuser}, \citenamefont {Heermeier}, \citenamefont {Lott}, \citenamefont {Reitzenstein}, \citenamefont {Hamerly},\ and\ \citenamefont {Englund}}]{Chen2023}%
  \BibitemOpen
  \bibfield  {author} {\bibinfo {author} {\bibfnamefont {Z.}~\bibnamefont {Chen}}, \bibinfo {author} {\bibfnamefont {A.}~\bibnamefont {Sludds}}, \bibinfo {author} {\bibfnamefont {R.}~\bibnamefont {Davis}}, \bibinfo {author} {\bibfnamefont {I.}~\bibnamefont {Christen}}, \bibinfo {author} {\bibfnamefont {L.}~\bibnamefont {Bernstein}}, \bibinfo {author} {\bibfnamefont {L.}~\bibnamefont {Ateshian}}, \bibinfo {author} {\bibfnamefont {T.}~\bibnamefont {Heuser}}, \bibinfo {author} {\bibfnamefont {N.}~\bibnamefont {Heermeier}}, \bibinfo {author} {\bibfnamefont {J.~A.}\ \bibnamefont {Lott}}, \bibinfo {author} {\bibfnamefont {S.}~\bibnamefont {Reitzenstein}}, \bibinfo {author} {\bibfnamefont {R.}~\bibnamefont {Hamerly}},\ and\ \bibinfo {author} {\bibfnamefont {D.}~\bibnamefont {Englund}},\ }\bibfield  {title} {\bibinfo {title} {Deep learning with coherent vcsel neural networks},\ }\href {https://doi.org/10.1038/s41566-023-01233-w} {\bibfield  {journal} {\bibinfo  {journal} {Nature Photonics}\ }\textbf {\bibinfo
  {volume} {17}},\ \bibinfo {pages} {723–730} (\bibinfo {year} {2023})}\BibitemShut {NoStop}%
\bibitem [{\citenamefont {Huo}\ \emph {et~al.}(2023)\citenamefont {Huo}, \citenamefont {Bao}, \citenamefont {Peng}, \citenamefont {Gao}, \citenamefont {Hua}, \citenamefont {Yang}, \citenamefont {Li}, \citenamefont {Wang},\ and\ \citenamefont {Yoon}}]{Huo2023}%
  \BibitemOpen
  \bibfield  {author} {\bibinfo {author} {\bibfnamefont {Y.}~\bibnamefont {Huo}}, \bibinfo {author} {\bibfnamefont {H.}~\bibnamefont {Bao}}, \bibinfo {author} {\bibfnamefont {Y.}~\bibnamefont {Peng}}, \bibinfo {author} {\bibfnamefont {C.}~\bibnamefont {Gao}}, \bibinfo {author} {\bibfnamefont {W.}~\bibnamefont {Hua}}, \bibinfo {author} {\bibfnamefont {Q.}~\bibnamefont {Yang}}, \bibinfo {author} {\bibfnamefont {H.}~\bibnamefont {Li}}, \bibinfo {author} {\bibfnamefont {R.}~\bibnamefont {Wang}},\ and\ \bibinfo {author} {\bibfnamefont {S.-E.}\ \bibnamefont {Yoon}},\ }\bibfield  {title} {\bibinfo {title} {Optical neural network via loose neuron array and functional learning},\ }\bibfield  {journal} {\bibinfo  {journal} {Nature Communications}\ }\textbf {\bibinfo {volume} {14}},\ \href {https://doi.org/10.1038/s41467-023-37390-3} {10.1038/s41467-023-37390-3} (\bibinfo {year} {2023})\BibitemShut {NoStop}%
\bibitem [{\citenamefont {Ou}\ \emph {et~al.}(2024)\citenamefont {Ou}, \citenamefont {Sludds}, \citenamefont {Hamerly}, \citenamefont {Zhang}, \citenamefont {Feng}, \citenamefont {Zhong}, \citenamefont {Wang}, \citenamefont {Englund}, \citenamefont {Yu},\ and\ \citenamefont {Chen}}]{Ou2024}%
  \BibitemOpen
  \bibfield  {author} {\bibinfo {author} {\bibfnamefont {S.}~\bibnamefont {Ou}}, \bibinfo {author} {\bibfnamefont {A.}~\bibnamefont {Sludds}}, \bibinfo {author} {\bibfnamefont {R.}~\bibnamefont {Hamerly}}, \bibinfo {author} {\bibfnamefont {K.}~\bibnamefont {Zhang}}, \bibinfo {author} {\bibfnamefont {H.}~\bibnamefont {Feng}}, \bibinfo {author} {\bibfnamefont {E.}~\bibnamefont {Zhong}}, \bibinfo {author} {\bibfnamefont {C.}~\bibnamefont {Wang}}, \bibinfo {author} {\bibfnamefont {D.}~\bibnamefont {Englund}}, \bibinfo {author} {\bibfnamefont {M.}~\bibnamefont {Yu}},\ and\ \bibinfo {author} {\bibfnamefont {Z.}~\bibnamefont {Chen}},\ }\href {https://arxiv.org/abs/2401.18050} {\bibinfo {title} {Hypermultiplexed integrated tensor optical processor}} (\bibinfo {year} {2024}),\ \Eprint {https://arxiv.org/abs/2401.18050} {arXiv:2401.18050 [cs.ET]} \BibitemShut {NoStop}%
\bibitem [{\citenamefont {Feldmann}\ \emph {et~al.}(2021)\citenamefont {Feldmann}, \citenamefont {Youngblood}, \citenamefont {Karpov}, \citenamefont {Gehring}, \citenamefont {Li}, \citenamefont {Stappers}, \citenamefont {Le~Gallo}, \citenamefont {Fu}, \citenamefont {Lukashchuk}, \citenamefont {Raja}, \citenamefont {Liu}, \citenamefont {Wright}, \citenamefont {Sebastian}, \citenamefont {Kippenberg}, \citenamefont {Pernice},\ and\ \citenamefont {Bhaskaran}}]{Feldmann2021}%
  \BibitemOpen
  \bibfield  {author} {\bibinfo {author} {\bibfnamefont {J.}~\bibnamefont {Feldmann}}, \bibinfo {author} {\bibfnamefont {N.}~\bibnamefont {Youngblood}}, \bibinfo {author} {\bibfnamefont {M.}~\bibnamefont {Karpov}}, \bibinfo {author} {\bibfnamefont {H.}~\bibnamefont {Gehring}}, \bibinfo {author} {\bibfnamefont {X.}~\bibnamefont {Li}}, \bibinfo {author} {\bibfnamefont {M.}~\bibnamefont {Stappers}}, \bibinfo {author} {\bibfnamefont {M.}~\bibnamefont {Le~Gallo}}, \bibinfo {author} {\bibfnamefont {X.}~\bibnamefont {Fu}}, \bibinfo {author} {\bibfnamefont {A.}~\bibnamefont {Lukashchuk}}, \bibinfo {author} {\bibfnamefont {A.~S.}\ \bibnamefont {Raja}}, \bibinfo {author} {\bibfnamefont {J.}~\bibnamefont {Liu}}, \bibinfo {author} {\bibfnamefont {C.~D.}\ \bibnamefont {Wright}}, \bibinfo {author} {\bibfnamefont {A.}~\bibnamefont {Sebastian}}, \bibinfo {author} {\bibfnamefont {T.~J.}\ \bibnamefont {Kippenberg}}, \bibinfo {author} {\bibfnamefont {W.~H.~P.}\ \bibnamefont {Pernice}},\ and\ \bibinfo {author} {\bibfnamefont
  {H.}~\bibnamefont {Bhaskaran}},\ }\bibfield  {title} {\bibinfo {title} {Parallel convolutional processing using an integrated photonic tensor core},\ }\href {https://doi.org/10.1038/s41586-020-03070-1} {\bibfield  {journal} {\bibinfo  {journal} {Nature}\ }\textbf {\bibinfo {volume} {589}},\ \bibinfo {pages} {52–58} (\bibinfo {year} {2021})}\BibitemShut {NoStop}%
\bibitem [{\citenamefont {Xu}\ \emph {et~al.}(2022)\citenamefont {Xu}, \citenamefont {Wang}, \citenamefont {Yi},\ and\ \citenamefont {Zou}}]{Xu2022}%
  \BibitemOpen
  \bibfield  {author} {\bibinfo {author} {\bibfnamefont {S.}~\bibnamefont {Xu}}, \bibinfo {author} {\bibfnamefont {J.}~\bibnamefont {Wang}}, \bibinfo {author} {\bibfnamefont {S.}~\bibnamefont {Yi}},\ and\ \bibinfo {author} {\bibfnamefont {W.}~\bibnamefont {Zou}},\ }\bibfield  {title} {\bibinfo {title} {High-order tensor flow processing using integrated photonic circuits},\ }\bibfield  {journal} {\bibinfo  {journal} {Nature Communications}\ }\textbf {\bibinfo {volume} {13}},\ \href {https://doi.org/10.1038/s41467-022-35723-2} {10.1038/s41467-022-35723-2} (\bibinfo {year} {2022})\BibitemShut {NoStop}%
\bibitem [{\citenamefont {Aggarwal}\ \emph {et~al.}(2023)\citenamefont {Aggarwal}, \citenamefont {Dong}, \citenamefont {Feldmann}, \citenamefont {Farmakidis}, \citenamefont {Pernice},\ and\ \citenamefont {Bhaskaran}}]{Aggarwal2023}%
  \BibitemOpen
  \bibfield  {author} {\bibinfo {author} {\bibfnamefont {S.}~\bibnamefont {Aggarwal}}, \bibinfo {author} {\bibfnamefont {B.}~\bibnamefont {Dong}}, \bibinfo {author} {\bibfnamefont {J.}~\bibnamefont {Feldmann}}, \bibinfo {author} {\bibfnamefont {N.}~\bibnamefont {Farmakidis}}, \bibinfo {author} {\bibfnamefont {W.~H.~P.}\ \bibnamefont {Pernice}},\ and\ \bibinfo {author} {\bibfnamefont {H.}~\bibnamefont {Bhaskaran}},\ }\bibfield  {title} {\bibinfo {title} {Reduced rank photonic computing accelerator},\ }\href {https://doi.org/10.1364/optica.485883} {\bibfield  {journal} {\bibinfo  {journal} {Optica}\ }\textbf {\bibinfo {volume} {10}},\ \bibinfo {pages} {1074} (\bibinfo {year} {2023})}\BibitemShut {NoStop}%
\bibitem [{\citenamefont {Dong}\ \emph {et~al.}(2023)\citenamefont {Dong}, \citenamefont {Aggarwal}, \citenamefont {Zhou}, \citenamefont {Ali}, \citenamefont {Farmakidis}, \citenamefont {Lee}, \citenamefont {He}, \citenamefont {Li}, \citenamefont {Kwong}, \citenamefont {Wright}, \citenamefont {Pernice},\ and\ \citenamefont {Bhaskaran}}]{Dong2023}%
  \BibitemOpen
  \bibfield  {author} {\bibinfo {author} {\bibfnamefont {B.}~\bibnamefont {Dong}}, \bibinfo {author} {\bibfnamefont {S.}~\bibnamefont {Aggarwal}}, \bibinfo {author} {\bibfnamefont {W.}~\bibnamefont {Zhou}}, \bibinfo {author} {\bibfnamefont {U.~E.}\ \bibnamefont {Ali}}, \bibinfo {author} {\bibfnamefont {N.}~\bibnamefont {Farmakidis}}, \bibinfo {author} {\bibfnamefont {J.~S.}\ \bibnamefont {Lee}}, \bibinfo {author} {\bibfnamefont {Y.}~\bibnamefont {He}}, \bibinfo {author} {\bibfnamefont {X.}~\bibnamefont {Li}}, \bibinfo {author} {\bibfnamefont {D.-L.}\ \bibnamefont {Kwong}}, \bibinfo {author} {\bibfnamefont {C.~D.}\ \bibnamefont {Wright}}, \bibinfo {author} {\bibfnamefont {W.~H.~P.}\ \bibnamefont {Pernice}},\ and\ \bibinfo {author} {\bibfnamefont {H.}~\bibnamefont {Bhaskaran}},\ }\bibfield  {title} {\bibinfo {title} {Higher-dimensional processing using a photonic tensor core with continuous-time data},\ }\href {https://doi.org/10.1038/s41566-023-01313-x} {\bibfield  {journal} {\bibinfo  {journal} {Nature
  Photonics}\ }\textbf {\bibinfo {volume} {17}},\ \bibinfo {pages} {1080–1088} (\bibinfo {year} {2023})}\BibitemShut {NoStop}%
\bibitem [{\citenamefont {Dong}\ \emph {et~al.}(2024)\citenamefont {Dong}, \citenamefont {Br\"{u}ckerhoff-Pl\"{u}ckelmann}, \citenamefont {Meyer}, \citenamefont {Dijkstra}, \citenamefont {Bente}, \citenamefont {Wendland}, \citenamefont {Varri}, \citenamefont {Aggarwal}, \citenamefont {Farmakidis}, \citenamefont {Wang}, \citenamefont {Yang}, \citenamefont {Lee}, \citenamefont {He}, \citenamefont {Gooskens}, \citenamefont {Kwong}, \citenamefont {Bienstman}, \citenamefont {Pernice},\ and\ \citenamefont {Bhaskaran}}]{Dong2024}%
  \BibitemOpen
  \bibfield  {author} {\bibinfo {author} {\bibfnamefont {B.}~\bibnamefont {Dong}}, \bibinfo {author} {\bibfnamefont {F.}~\bibnamefont {Br\"{u}ckerhoff-Pl\"{u}ckelmann}}, \bibinfo {author} {\bibfnamefont {L.}~\bibnamefont {Meyer}}, \bibinfo {author} {\bibfnamefont {J.}~\bibnamefont {Dijkstra}}, \bibinfo {author} {\bibfnamefont {I.}~\bibnamefont {Bente}}, \bibinfo {author} {\bibfnamefont {D.}~\bibnamefont {Wendland}}, \bibinfo {author} {\bibfnamefont {A.}~\bibnamefont {Varri}}, \bibinfo {author} {\bibfnamefont {S.}~\bibnamefont {Aggarwal}}, \bibinfo {author} {\bibfnamefont {N.}~\bibnamefont {Farmakidis}}, \bibinfo {author} {\bibfnamefont {M.}~\bibnamefont {Wang}}, \bibinfo {author} {\bibfnamefont {G.}~\bibnamefont {Yang}}, \bibinfo {author} {\bibfnamefont {J.~S.}\ \bibnamefont {Lee}}, \bibinfo {author} {\bibfnamefont {Y.}~\bibnamefont {He}}, \bibinfo {author} {\bibfnamefont {E.}~\bibnamefont {Gooskens}}, \bibinfo {author} {\bibfnamefont {D.-L.}\ \bibnamefont {Kwong}}, \bibinfo {author} {\bibfnamefont
  {P.}~\bibnamefont {Bienstman}}, \bibinfo {author} {\bibfnamefont {W.~H.~P.}\ \bibnamefont {Pernice}},\ and\ \bibinfo {author} {\bibfnamefont {H.}~\bibnamefont {Bhaskaran}},\ }\bibfield  {title} {\bibinfo {title} {Partial coherence enhances parallelized photonic computing},\ }\href {https://doi.org/10.1038/s41586-024-07590-y} {\bibfield  {journal} {\bibinfo  {journal} {Nature}\ }\textbf {\bibinfo {volume} {632}},\ \bibinfo {pages} {55–62} (\bibinfo {year} {2024})}\BibitemShut {NoStop}%
\bibitem [{\citenamefont {Moralis-Pegios}\ \emph {et~al.}(2024)\citenamefont {Moralis-Pegios}, \citenamefont {Giamougiannis}, \citenamefont {Tsakyridis}, \citenamefont {Lazovsky},\ and\ \citenamefont {Pleros}}]{MoralisPegios2024}%
  \BibitemOpen
  \bibfield  {author} {\bibinfo {author} {\bibfnamefont {M.}~\bibnamefont {Moralis-Pegios}}, \bibinfo {author} {\bibfnamefont {G.}~\bibnamefont {Giamougiannis}}, \bibinfo {author} {\bibfnamefont {A.}~\bibnamefont {Tsakyridis}}, \bibinfo {author} {\bibfnamefont {D.}~\bibnamefont {Lazovsky}},\ and\ \bibinfo {author} {\bibfnamefont {N.}~\bibnamefont {Pleros}},\ }\bibfield  {title} {\bibinfo {title} {Perfect linear optics using silicon photonics},\ }\bibfield  {journal} {\bibinfo  {journal} {Nature Communications}\ }\textbf {\bibinfo {volume} {15}},\ \href {https://doi.org/10.1038/s41467-024-49768-y} {10.1038/s41467-024-49768-y} (\bibinfo {year} {2024})\BibitemShut {NoStop}%
\bibitem [{\citenamefont {Reck}\ \emph {et~al.}(1994)\citenamefont {Reck}, \citenamefont {Zeilinger}, \citenamefont {Bernstein},\ and\ \citenamefont {Bertani}}]{Reck1994}%
  \BibitemOpen
  \bibfield  {author} {\bibinfo {author} {\bibfnamefont {M.}~\bibnamefont {Reck}}, \bibinfo {author} {\bibfnamefont {A.}~\bibnamefont {Zeilinger}}, \bibinfo {author} {\bibfnamefont {H.~J.}\ \bibnamefont {Bernstein}},\ and\ \bibinfo {author} {\bibfnamefont {P.}~\bibnamefont {Bertani}},\ }\bibfield  {title} {\bibinfo {title} {Experimental realization of any discrete unitary operator},\ }\href {https://doi.org/10.1103/PhysRevLett.73.58} {\bibfield  {journal} {\bibinfo  {journal} {Phys. Rev. Lett.}\ }\textbf {\bibinfo {volume} {73}},\ \bibinfo {pages} {58} (\bibinfo {year} {1994})}\BibitemShut {NoStop}%
\bibitem [{\citenamefont {Carolan}\ \emph {et~al.}(2015)\citenamefont {Carolan}, \citenamefont {Harrold}, \citenamefont {Sparrow}, \citenamefont {Martín-López}, \citenamefont {Russell}, \citenamefont {Silverstone}, \citenamefont {Shadbolt}, \citenamefont {Matsuda}, \citenamefont {Oguma}, \citenamefont {Itoh}, \citenamefont {Marshall}, \citenamefont {Thompson}, \citenamefont {Matthews}, \citenamefont {Hashimoto}, \citenamefont {O’Brien},\ and\ \citenamefont {Laing}}]{Carolan2015}%
  \BibitemOpen
  \bibfield  {author} {\bibinfo {author} {\bibfnamefont {J.}~\bibnamefont {Carolan}}, \bibinfo {author} {\bibfnamefont {C.}~\bibnamefont {Harrold}}, \bibinfo {author} {\bibfnamefont {C.}~\bibnamefont {Sparrow}}, \bibinfo {author} {\bibfnamefont {E.}~\bibnamefont {Martín-López}}, \bibinfo {author} {\bibfnamefont {N.~J.}\ \bibnamefont {Russell}}, \bibinfo {author} {\bibfnamefont {J.~W.}\ \bibnamefont {Silverstone}}, \bibinfo {author} {\bibfnamefont {P.~J.}\ \bibnamefont {Shadbolt}}, \bibinfo {author} {\bibfnamefont {N.}~\bibnamefont {Matsuda}}, \bibinfo {author} {\bibfnamefont {M.}~\bibnamefont {Oguma}}, \bibinfo {author} {\bibfnamefont {M.}~\bibnamefont {Itoh}}, \bibinfo {author} {\bibfnamefont {G.~D.}\ \bibnamefont {Marshall}}, \bibinfo {author} {\bibfnamefont {M.~G.}\ \bibnamefont {Thompson}}, \bibinfo {author} {\bibfnamefont {J.~C.~F.}\ \bibnamefont {Matthews}}, \bibinfo {author} {\bibfnamefont {T.}~\bibnamefont {Hashimoto}}, \bibinfo {author} {\bibfnamefont {J.~L.}\ \bibnamefont {O’Brien}},\ and\
  \bibinfo {author} {\bibfnamefont {A.}~\bibnamefont {Laing}},\ }\bibfield  {title} {\bibinfo {title} {Universal linear optics},\ }\href {https://doi.org/10.1126/science.aab3642} {\bibfield  {journal} {\bibinfo  {journal} {Science}\ }\textbf {\bibinfo {volume} {349}},\ \bibinfo {pages} {711–716} (\bibinfo {year} {2015})}\BibitemShut {NoStop}%
\bibitem [{\citenamefont {Clements}\ \emph {et~al.}(2016)\citenamefont {Clements}, \citenamefont {Humphreys}, \citenamefont {Metcalf}, \citenamefont {Kolthammer},\ and\ \citenamefont {Walsmley}}]{Clements2016}%
  \BibitemOpen
  \bibfield  {author} {\bibinfo {author} {\bibfnamefont {W.~R.}\ \bibnamefont {Clements}}, \bibinfo {author} {\bibfnamefont {P.~C.}\ \bibnamefont {Humphreys}}, \bibinfo {author} {\bibfnamefont {B.~J.}\ \bibnamefont {Metcalf}}, \bibinfo {author} {\bibfnamefont {W.~S.}\ \bibnamefont {Kolthammer}},\ and\ \bibinfo {author} {\bibfnamefont {I.~A.}\ \bibnamefont {Walsmley}},\ }\bibfield  {title} {\bibinfo {title} {Optimal design for universal multiport interferometers},\ }\href {https://doi.org/10.1364/optica.3.001460} {\bibfield  {journal} {\bibinfo  {journal} {Optica}\ }\textbf {\bibinfo {volume} {3}},\ \bibinfo {pages} {1460} (\bibinfo {year} {2016})}\BibitemShut {NoStop}%
\bibitem [{\citenamefont {Shen}\ \emph {et~al.}(2017)\citenamefont {Shen}, \citenamefont {Harris}, \citenamefont {Skirlo}, \citenamefont {Prabhu}, \citenamefont {Baehr-Jones}, \citenamefont {Hochberg}, \citenamefont {Sun}, \citenamefont {Zhao}, \citenamefont {Larochelle}, \citenamefont {Englund},\ and\ \citenamefont {Soljačić}}]{Shen2017}%
  \BibitemOpen
  \bibfield  {author} {\bibinfo {author} {\bibfnamefont {Y.}~\bibnamefont {Shen}}, \bibinfo {author} {\bibfnamefont {N.~C.}\ \bibnamefont {Harris}}, \bibinfo {author} {\bibfnamefont {S.}~\bibnamefont {Skirlo}}, \bibinfo {author} {\bibfnamefont {M.}~\bibnamefont {Prabhu}}, \bibinfo {author} {\bibfnamefont {T.}~\bibnamefont {Baehr-Jones}}, \bibinfo {author} {\bibfnamefont {M.}~\bibnamefont {Hochberg}}, \bibinfo {author} {\bibfnamefont {X.}~\bibnamefont {Sun}}, \bibinfo {author} {\bibfnamefont {S.}~\bibnamefont {Zhao}}, \bibinfo {author} {\bibfnamefont {H.}~\bibnamefont {Larochelle}}, \bibinfo {author} {\bibfnamefont {D.}~\bibnamefont {Englund}},\ and\ \bibinfo {author} {\bibfnamefont {M.}~\bibnamefont {Soljačić}},\ }\bibfield  {title} {\bibinfo {title} {Deep learning with coherent nanophotonic circuits},\ }\href {https://doi.org/10.1038/nphoton.2017.93} {\bibfield  {journal} {\bibinfo  {journal} {Nature Photonics}\ }\textbf {\bibinfo {volume} {11}},\ \bibinfo {pages} {441–446} (\bibinfo {year}
  {2017})}\BibitemShut {NoStop}%
\bibitem [{\citenamefont {Pérez}\ \emph {et~al.}(2017)\citenamefont {Pérez}, \citenamefont {Gasulla}, \citenamefont {Crudgington}, \citenamefont {Thomson}, \citenamefont {Khokhar}, \citenamefont {Li}, \citenamefont {Cao}, \citenamefont {Mashanovich},\ and\ \citenamefont {Capmany}}]{Prez2017}%
  \BibitemOpen
  \bibfield  {author} {\bibinfo {author} {\bibfnamefont {D.}~\bibnamefont {Pérez}}, \bibinfo {author} {\bibfnamefont {I.}~\bibnamefont {Gasulla}}, \bibinfo {author} {\bibfnamefont {L.}~\bibnamefont {Crudgington}}, \bibinfo {author} {\bibfnamefont {D.~J.}\ \bibnamefont {Thomson}}, \bibinfo {author} {\bibfnamefont {A.~Z.}\ \bibnamefont {Khokhar}}, \bibinfo {author} {\bibfnamefont {K.}~\bibnamefont {Li}}, \bibinfo {author} {\bibfnamefont {W.}~\bibnamefont {Cao}}, \bibinfo {author} {\bibfnamefont {G.~Z.}\ \bibnamefont {Mashanovich}},\ and\ \bibinfo {author} {\bibfnamefont {J.}~\bibnamefont {Capmany}},\ }\bibfield  {title} {\bibinfo {title} {Multipurpose silicon photonics signal processor core},\ }\bibfield  {journal} {\bibinfo  {journal} {Nature Communications}\ }\textbf {\bibinfo {volume} {8}},\ \href {https://doi.org/10.1038/s41467-017-00714-1} {10.1038/s41467-017-00714-1} (\bibinfo {year} {2017})\BibitemShut {NoStop}%
\bibitem [{\citenamefont {Hughes}\ \emph {et~al.}(2018{\natexlab{a}})\citenamefont {Hughes}, \citenamefont {Minkov}, \citenamefont {Shi},\ and\ \citenamefont {Fan}}]{Hughes2018optica}%
  \BibitemOpen
  \bibfield  {author} {\bibinfo {author} {\bibfnamefont {T.~W.}\ \bibnamefont {Hughes}}, \bibinfo {author} {\bibfnamefont {M.}~\bibnamefont {Minkov}}, \bibinfo {author} {\bibfnamefont {Y.}~\bibnamefont {Shi}},\ and\ \bibinfo {author} {\bibfnamefont {S.}~\bibnamefont {Fan}},\ }\bibfield  {title} {\bibinfo {title} {Training of photonic neural networks through in situ backpropagation and gradient measurement},\ }\href {https://doi.org/10.1364/optica.5.000864} {\bibfield  {journal} {\bibinfo  {journal} {Optica}\ }\textbf {\bibinfo {volume} {5}},\ \bibinfo {pages} {864} (\bibinfo {year} {2018}{\natexlab{a}})}\BibitemShut {NoStop}%
\bibitem [{\citenamefont {Khoram}\ \emph {et~al.}(2019)\citenamefont {Khoram}, \citenamefont {Chen}, \citenamefont {Liu}, \citenamefont {Ying}, \citenamefont {Wang}, \citenamefont {Yuan},\ and\ \citenamefont {Yu}}]{Khoram2019}%
  \BibitemOpen
  \bibfield  {author} {\bibinfo {author} {\bibfnamefont {E.}~\bibnamefont {Khoram}}, \bibinfo {author} {\bibfnamefont {A.}~\bibnamefont {Chen}}, \bibinfo {author} {\bibfnamefont {D.}~\bibnamefont {Liu}}, \bibinfo {author} {\bibfnamefont {L.}~\bibnamefont {Ying}}, \bibinfo {author} {\bibfnamefont {Q.}~\bibnamefont {Wang}}, \bibinfo {author} {\bibfnamefont {M.}~\bibnamefont {Yuan}},\ and\ \bibinfo {author} {\bibfnamefont {Z.}~\bibnamefont {Yu}},\ }\bibfield  {title} {\bibinfo {title} {Nanophotonic media for artificial neural inference},\ }\href {https://doi.org/10.1364/prj.7.000823} {\bibfield  {journal} {\bibinfo  {journal} {Photonics Research}\ }\textbf {\bibinfo {volume} {7}},\ \bibinfo {pages} {823} (\bibinfo {year} {2019})}\BibitemShut {NoStop}%
\bibitem [{\citenamefont {Camacho}\ \emph {et~al.}(2021)\citenamefont {Camacho}, \citenamefont {Edwards},\ and\ \citenamefont {Engheta}}]{Camacho2021}%
  \BibitemOpen
  \bibfield  {author} {\bibinfo {author} {\bibfnamefont {M.}~\bibnamefont {Camacho}}, \bibinfo {author} {\bibfnamefont {B.}~\bibnamefont {Edwards}},\ and\ \bibinfo {author} {\bibfnamefont {N.}~\bibnamefont {Engheta}},\ }\bibfield  {title} {\bibinfo {title} {A single inverse-designed photonic structure that performs parallel computing},\ }\bibfield  {journal} {\bibinfo  {journal} {Nature Communications}\ }\textbf {\bibinfo {volume} {12}},\ \href {https://doi.org/10.1038/s41467-021-21664-9} {10.1038/s41467-021-21664-9} (\bibinfo {year} {2021})\BibitemShut {NoStop}%
\bibitem [{\citenamefont {Nikkhah}\ \emph {et~al.}(2024)\citenamefont {Nikkhah}, \citenamefont {Pirmoradi}, \citenamefont {Ashtiani}, \citenamefont {Edwards}, \citenamefont {Aflatouni},\ and\ \citenamefont {Engheta}}]{Nikkhah2024}%
  \BibitemOpen
  \bibfield  {author} {\bibinfo {author} {\bibfnamefont {V.}~\bibnamefont {Nikkhah}}, \bibinfo {author} {\bibfnamefont {A.}~\bibnamefont {Pirmoradi}}, \bibinfo {author} {\bibfnamefont {F.}~\bibnamefont {Ashtiani}}, \bibinfo {author} {\bibfnamefont {B.}~\bibnamefont {Edwards}}, \bibinfo {author} {\bibfnamefont {F.}~\bibnamefont {Aflatouni}},\ and\ \bibinfo {author} {\bibfnamefont {N.}~\bibnamefont {Engheta}},\ }\bibfield  {title} {\bibinfo {title} {Inverse-designed low-index-contrast structures on a silicon photonics platform for vector–matrix multiplication},\ }\href {https://doi.org/10.1038/s41566-024-01394-2} {\bibfield  {journal} {\bibinfo  {journal} {Nature Photonics}\ }\textbf {\bibinfo {volume} {18}},\ \bibinfo {pages} {501–508} (\bibinfo {year} {2024})}\BibitemShut {NoStop}%
\bibitem [{\citenamefont {Youngblood}(2023)}]{Youngblood2023}%
  \BibitemOpen
  \bibfield  {author} {\bibinfo {author} {\bibfnamefont {N.}~\bibnamefont {Youngblood}},\ }\bibfield  {title} {\bibinfo {title} {Coherent photonic crossbar arrays for large-scale matrix-matrix multiplication},\ }\href {https://doi.org/10.1109/JSTQE.2022.3171167} {\bibfield  {journal} {\bibinfo  {journal} {IEEE Journal of Selected Topics in Quantum Electronics}\ }\textbf {\bibinfo {volume} {29}},\ \bibinfo {pages} {1} (\bibinfo {year} {2023})}\BibitemShut {NoStop}%
\bibitem [{\citenamefont {Kari}\ \emph {et~al.}(2024)\citenamefont {Kari}, \citenamefont {Nobile}, \citenamefont {Pantin}, \citenamefont {Shah},\ and\ \citenamefont {Youngblood}}]{RahimiKari2024}%
  \BibitemOpen
  \bibfield  {author} {\bibinfo {author} {\bibfnamefont {S.~R.}\ \bibnamefont {Kari}}, \bibinfo {author} {\bibfnamefont {N.~A.}\ \bibnamefont {Nobile}}, \bibinfo {author} {\bibfnamefont {D.}~\bibnamefont {Pantin}}, \bibinfo {author} {\bibfnamefont {V.}~\bibnamefont {Shah}},\ and\ \bibinfo {author} {\bibfnamefont {N.}~\bibnamefont {Youngblood}},\ }\bibfield  {title} {\bibinfo {title} {Realization of an integrated coherent photonic platform for scalable matrix operations},\ }\href {https://doi.org/10.1364/OPTICA.507525} {\bibfield  {journal} {\bibinfo  {journal} {Optica}\ }\textbf {\bibinfo {volume} {11}},\ \bibinfo {pages} {542} (\bibinfo {year} {2024})}\BibitemShut {NoStop}%
\bibitem [{\citenamefont {Lalau-Keraly}\ \emph {et~al.}(2013)\citenamefont {Lalau-Keraly}, \citenamefont {Bhargava}, \citenamefont {Miller},\ and\ \citenamefont {Yablonovitch}}]{LalauKeraly2013}%
  \BibitemOpen
  \bibfield  {author} {\bibinfo {author} {\bibfnamefont {C.~M.}\ \bibnamefont {Lalau-Keraly}}, \bibinfo {author} {\bibfnamefont {S.}~\bibnamefont {Bhargava}}, \bibinfo {author} {\bibfnamefont {O.~D.}\ \bibnamefont {Miller}},\ and\ \bibinfo {author} {\bibfnamefont {E.}~\bibnamefont {Yablonovitch}},\ }\bibfield  {title} {\bibinfo {title} {Adjoint shape optimization applied to electromagnetic design},\ }\href {https://doi.org/10.1364/oe.21.021693} {\bibfield  {journal} {\bibinfo  {journal} {Optics Express}\ }\textbf {\bibinfo {volume} {21}},\ \bibinfo {pages} {21693} (\bibinfo {year} {2013})}\BibitemShut {NoStop}%
\bibitem [{\citenamefont {Hughes}\ \emph {et~al.}(2018{\natexlab{b}})\citenamefont {Hughes}, \citenamefont {Minkov}, \citenamefont {Williamson},\ and\ \citenamefont {Fan}}]{Hughes2018}%
  \BibitemOpen
  \bibfield  {author} {\bibinfo {author} {\bibfnamefont {T.~W.}\ \bibnamefont {Hughes}}, \bibinfo {author} {\bibfnamefont {M.}~\bibnamefont {Minkov}}, \bibinfo {author} {\bibfnamefont {I.~A.~D.}\ \bibnamefont {Williamson}},\ and\ \bibinfo {author} {\bibfnamefont {S.}~\bibnamefont {Fan}},\ }\bibfield  {title} {\bibinfo {title} {Adjoint method and inverse design for nonlinear nanophotonic devices},\ }\href {https://doi.org/10.1021/acsphotonics.8b01522} {\bibfield  {journal} {\bibinfo  {journal} {ACS Photonics}\ }\textbf {\bibinfo {volume} {5}},\ \bibinfo {pages} {4781–4787} (\bibinfo {year} {2018}{\natexlab{b}})}\BibitemShut {NoStop}%
\bibitem [{\citenamefont {Hammond}\ \emph {et~al.}(2022)\citenamefont {Hammond}, \citenamefont {Slaby}, \citenamefont {Probst},\ and\ \citenamefont {Ralph}}]{Hammond2022}%
  \BibitemOpen
  \bibfield  {author} {\bibinfo {author} {\bibfnamefont {A.~M.}\ \bibnamefont {Hammond}}, \bibinfo {author} {\bibfnamefont {J.~B.}\ \bibnamefont {Slaby}}, \bibinfo {author} {\bibfnamefont {M.~J.}\ \bibnamefont {Probst}},\ and\ \bibinfo {author} {\bibfnamefont {S.~E.}\ \bibnamefont {Ralph}},\ }\bibfield  {title} {\bibinfo {title} {Multi-layer inverse design of vertical grating couplers for high-density, commercial foundry interconnects},\ }\href {https://doi.org/10.1364/OE.466015} {\bibfield  {journal} {\bibinfo  {journal} {Opt. Express}\ }\textbf {\bibinfo {volume} {30}},\ \bibinfo {pages} {31058} (\bibinfo {year} {2022})}\BibitemShut {NoStop}%
\bibitem [{\citenamefont {Hammond}\ \emph {et~al.}(2023)\citenamefont {Hammond}, \citenamefont {Slaby}, \citenamefont {Probst},\ and\ \citenamefont {Ralph}}]{Hammond2023}%
  \BibitemOpen
  \bibfield  {author} {\bibinfo {author} {\bibfnamefont {A.~M.}\ \bibnamefont {Hammond}}, \bibinfo {author} {\bibfnamefont {J.~B.}\ \bibnamefont {Slaby}}, \bibinfo {author} {\bibfnamefont {M.~J.}\ \bibnamefont {Probst}},\ and\ \bibinfo {author} {\bibfnamefont {S.~E.}\ \bibnamefont {Ralph}},\ }\bibfield  {title} {\bibinfo {title} {Phase-injected topology optimization for scalable and interferometrically robust photonic integrated circuits},\ }\href {https://doi.org/10.1021/acsphotonics.2c01016} {\bibfield  {journal} {\bibinfo  {journal} {ACS Photonics}\ }\textbf {\bibinfo {volume} {10}},\ \bibinfo {pages} {808} (\bibinfo {year} {2023})},\ \Eprint {https://arxiv.org/abs/https://doi.org/10.1021/acsphotonics.2c01016} {https://doi.org/10.1021/acsphotonics.2c01016} \BibitemShut {NoStop}%
\bibitem [{\citenamefont {Wu}\ \emph {et~al.}(2023)\citenamefont {Wu}, \citenamefont {Menarini}, \citenamefont {Gao},\ and\ \citenamefont {Feng}}]{Wu2023}%
  \BibitemOpen
  \bibfield  {author} {\bibinfo {author} {\bibfnamefont {T.}~\bibnamefont {Wu}}, \bibinfo {author} {\bibfnamefont {M.}~\bibnamefont {Menarini}}, \bibinfo {author} {\bibfnamefont {Z.}~\bibnamefont {Gao}},\ and\ \bibinfo {author} {\bibfnamefont {L.}~\bibnamefont {Feng}},\ }\bibfield  {title} {\bibinfo {title} {Lithography-free reconfigurable integrated photonic processor},\ }\href {https://doi.org/10.1038/s41566-023-01205-0} {\bibfield  {journal} {\bibinfo  {journal} {Nature Photonics}\ }\textbf {\bibinfo {volume} {17}},\ \bibinfo {pages} {710–716} (\bibinfo {year} {2023})}\BibitemShut {NoStop}%
\bibitem [{\citenamefont {FlexCompute}()}]{Tidy3d}%
  \BibitemOpen
  \bibfield  {author} {\bibinfo {author} {\bibnamefont {FlexCompute}},\ }\href@noop {} {\bibinfo {title} {Python-driven {FDTD} software: Tidy3{D}.}},\ \bibinfo {howpublished} {\url{https://www.flexcompute.com/tidy3d/solver/}},\ \bibinfo {note} {accessed: \today}\BibitemShut {NoStop}%
\bibitem [{\citenamefont {Minkov}\ \emph {et~al.}(2024)\citenamefont {Minkov}, \citenamefont {Sun}, \citenamefont {Lee}, \citenamefont {Yu},\ and\ \citenamefont {Fan}}]{Minkov2024}%
  \BibitemOpen
  \bibfield  {author} {\bibinfo {author} {\bibfnamefont {M.}~\bibnamefont {Minkov}}, \bibinfo {author} {\bibfnamefont {P.}~\bibnamefont {Sun}}, \bibinfo {author} {\bibfnamefont {B.}~\bibnamefont {Lee}}, \bibinfo {author} {\bibfnamefont {Z.}~\bibnamefont {Yu}},\ and\ \bibinfo {author} {\bibfnamefont {S.}~\bibnamefont {Fan}},\ }\bibfield  {title} {\bibinfo {title} {Gpu-accelerated photonic simulations},\ }\href {https://doi.org/10.1364/OPN.35.9.000044} {\bibfield  {journal} {\bibinfo  {journal} {Opt. Photon. News}\ }\textbf {\bibinfo {volume} {35}},\ \bibinfo {pages} {44} (\bibinfo {year} {2024})}\BibitemShut {NoStop}%
\bibitem [{\citenamefont {Fano}(1950)}]{Fano1950}%
  \BibitemOpen
  \bibfield  {author} {\bibinfo {author} {\bibfnamefont {R.}~\bibnamefont {Fano}},\ }\bibfield  {title} {\bibinfo {title} {Theoretical limitations on the broadband matching of arbitrary impedances},\ }\href {https://doi.org/https://doi.org/10.1016/0016-0032(50)90006-8} {\bibfield  {journal} {\bibinfo  {journal} {Journal of the Franklin Institute}\ }\textbf {\bibinfo {volume} {249}},\ \bibinfo {pages} {57} (\bibinfo {year} {1950})}\BibitemShut {NoStop}%
\bibitem [{\citenamefont {Tchernycheva}\ \emph {et~al.}(2014)\citenamefont {Tchernycheva}, \citenamefont {Messanvi}, \citenamefont {de~Luna~Bugallo}, \citenamefont {Jacopin}, \citenamefont {Lavenus}, \citenamefont {Rigutti}, \citenamefont {Zhang}, \citenamefont {Halioua}, \citenamefont {Julien}, \citenamefont {Eymery},\ and\ \citenamefont {Durand}}]{Tchernycheva2014}%
  \BibitemOpen
  \bibfield  {author} {\bibinfo {author} {\bibfnamefont {M.}~\bibnamefont {Tchernycheva}}, \bibinfo {author} {\bibfnamefont {A.}~\bibnamefont {Messanvi}}, \bibinfo {author} {\bibfnamefont {A.}~\bibnamefont {de~Luna~Bugallo}}, \bibinfo {author} {\bibfnamefont {G.}~\bibnamefont {Jacopin}}, \bibinfo {author} {\bibfnamefont {P.}~\bibnamefont {Lavenus}}, \bibinfo {author} {\bibfnamefont {L.}~\bibnamefont {Rigutti}}, \bibinfo {author} {\bibfnamefont {H.}~\bibnamefont {Zhang}}, \bibinfo {author} {\bibfnamefont {Y.}~\bibnamefont {Halioua}}, \bibinfo {author} {\bibfnamefont {F.~H.}\ \bibnamefont {Julien}}, \bibinfo {author} {\bibfnamefont {J.}~\bibnamefont {Eymery}},\ and\ \bibinfo {author} {\bibfnamefont {C.}~\bibnamefont {Durand}},\ }\bibfield  {title} {\bibinfo {title} {Integrated photonic platform based on ingan/gan nanowire emitters and detectors},\ }\href {https://doi.org/10.1021/nl501124s} {\bibfield  {journal} {\bibinfo  {journal} {Nano Letters}\ }\textbf {\bibinfo {volume} {14}},\ \bibinfo {pages} {3515}
  (\bibinfo {year} {2014})},\ \bibinfo {note} {pMID: 24837282},\ \Eprint {https://arxiv.org/abs/https://doi.org/10.1021/nl501124s} {https://doi.org/10.1021/nl501124s} \BibitemShut {NoStop}%
\bibitem [{\citenamefont {Gan}\ \emph {et~al.}(2013)\citenamefont {Gan}, \citenamefont {Shiue}, \citenamefont {Gao}, \citenamefont {Meric}, \citenamefont {Heinz}, \citenamefont {Shepard}, \citenamefont {Hone}, \citenamefont {Assefa},\ and\ \citenamefont {Englund}}]{Gan2013}%
  \BibitemOpen
  \bibfield  {author} {\bibinfo {author} {\bibfnamefont {X.}~\bibnamefont {Gan}}, \bibinfo {author} {\bibfnamefont {R.-J.}\ \bibnamefont {Shiue}}, \bibinfo {author} {\bibfnamefont {Y.}~\bibnamefont {Gao}}, \bibinfo {author} {\bibfnamefont {I.}~\bibnamefont {Meric}}, \bibinfo {author} {\bibfnamefont {T.~F.}\ \bibnamefont {Heinz}}, \bibinfo {author} {\bibfnamefont {K.}~\bibnamefont {Shepard}}, \bibinfo {author} {\bibfnamefont {J.}~\bibnamefont {Hone}}, \bibinfo {author} {\bibfnamefont {S.}~\bibnamefont {Assefa}},\ and\ \bibinfo {author} {\bibfnamefont {D.}~\bibnamefont {Englund}},\ }\bibfield  {title} {\bibinfo {title} {Chip-integrated ultrafast graphene photodetector with high responsivity},\ }\href {https://doi.org/10.1038/nphoton.2013.253} {\bibfield  {journal} {\bibinfo  {journal} {Nature Photonics}\ }\textbf {\bibinfo {volume} {7}},\ \bibinfo {pages} {883–887} (\bibinfo {year} {2013})}\BibitemShut {NoStop}%
\end{thebibliography}%


\begin{thebibliography}{2}%
\makeatletter
\providecommand \@ifxundefined [1]{%
 \@ifx{#1\undefined}
}%
\providecommand \@ifnum [1]{%
 \ifnum #1\expandafter \@firstoftwo
 \else \expandafter \@secondoftwo
 \fi
}%
\providecommand \@ifx [1]{%
 \ifx #1\expandafter \@firstoftwo
 \else \expandafter \@secondoftwo
 \fi
}%
\providecommand \natexlab [1]{#1}%
\providecommand \enquote  [1]{``#1''}%
\providecommand \bibnamefont  [1]{#1}%
\providecommand \bibfnamefont [1]{#1}%
\providecommand \citenamefont [1]{#1}%
\providecommand \href@noop [0]{\@secondoftwo}%
\providecommand \href [0]{\begingroup \@sanitize@url \@href}%
\providecommand \@href[1]{\@@startlink{#1}\@@href}%
\providecommand \@@href[1]{\endgroup#1\@@endlink}%
\providecommand \@sanitize@url [0]{\catcode `\\12\catcode `\$12\catcode `\&12\catcode `\#12\catcode `\^12\catcode `\_12\catcode `\%12\relax}%
\providecommand \@@startlink[1]{}%
\providecommand \@@endlink[0]{}%
\providecommand \url  [0]{\begingroup\@sanitize@url \@url }%
\providecommand \@url [1]{\endgroup\@href {#1}{\urlprefix }}%
\providecommand \urlprefix  [0]{URL }%
\providecommand \Eprint [0]{\href }%
\providecommand \doibase [0]{https://doi.org/}%
\providecommand \selectlanguage [0]{\@gobble}%
\providecommand \bibinfo  [0]{\@secondoftwo}%
\providecommand \bibfield  [0]{\@secondoftwo}%
\providecommand \translation [1]{[#1]}%
\providecommand \BibitemOpen [0]{}%
\providecommand \bibitemStop [0]{}%
\providecommand \bibitemNoStop [0]{.\EOS\space}%
\providecommand \EOS [0]{\spacefactor3000\relax}%
\providecommand \BibitemShut  [1]{\csname bibitem#1\endcsname}%
\let\auto@bib@innerbib\@empty
\bibitem [{\citenamefont {Youngblood}(2023)}]{Youngblood2023}%
  \BibitemOpen
  \bibfield  {author} {\bibinfo {author} {\bibfnamefont {N.}~\bibnamefont {Youngblood}},\ }\bibfield  {title} {\bibinfo {title} {Coherent photonic crossbar arrays for large-scale matrix-matrix multiplication},\ }\href {https://doi.org/10.1109/JSTQE.2022.3171167} {\bibfield  {journal} {\bibinfo  {journal} {IEEE Journal of Selected Topics in Quantum Electronics}\ }\textbf {\bibinfo {volume} {29}},\ \bibinfo {pages} {1} (\bibinfo {year} {2023})}\BibitemShut {NoStop}%
\bibitem [{\citenamefont {Palik}(1998)}]{material}%
  \BibitemOpen
  \bibfield  {author} {\bibinfo {author} {\bibfnamefont {E.~D.}\ \bibnamefont {Palik}},\ }\href@noop {} {\emph {\bibinfo {title} {Handbook of optical constants of solids}}}\ (\bibinfo  {publisher} {Academic Press},\ \bibinfo {address} {San Diego, CA},\ \bibinfo {year} {1998})\BibitemShut {NoStop}%
\end{thebibliography}%

\end{document}



\title{Supplementary information for ``Photonic systolic array for all-optical matrix-matrix multiplication"}%

\author{Jungmin Kim}
\email{jkim2325@wisc.edu}
\affiliation{Department of Electrical and Computer Engineering, University of Wisconsin-Madison, Madison, WI 53706, USA}
\affiliation{These authors contributed equally.}

\author{Qingyi Zhou}
\email{qzhou75@wisc.edu}
\affiliation{Department of Electrical and Computer Engineering, University of Wisconsin-Madison, Madison, WI 53706, USA}
\affiliation{These authors contributed equally.}

\author{Zongfu Yu}
\affiliation{Department of Electrical and Computer Engineering, University of Wisconsin-Madison, Madison, WI 53706, USA}

\date{\today}

\maketitle

\section{Loss-compensated fan-out design}


Consider an ideal $N \times N$ photonic systolic array (PSA) with submodules in the main text, where $1 \leq m, n \leq N$ are waveguide indices. In this ideal case without any insertion loss, the input power is uniformly distributed across the MAC units, with each unit receiving a fraction of $N^{-1}$. Thus, the ideal power split ratio at each branch of the $(m, n)$ MAC unit should be $m^{-1}$ and $n^{-1}$, respectively. Practically, however, there exist small but non-negligible losses, which must be accounted for when setting the target amplitude of the submodules, as described in Ref. \cite{Youngblood2023}.

We define the insertion losses for the waveguide crossing and the branches indexed by $n$ as: \begin{align} 
\alpha_\mathrm{cross} &\equiv |S_{13}|^2 \leq 1, \nonumber \\
\alpha_n &\equiv |S_{13}|^2 + |S_{23}|^2 \leq 1, \end{align} 
leading to a total loss $\alpha = \alpha_\mathrm{cross}\alpha_n$ for a signal passing through a MAC unit. Let $\rho_n \equiv |S_{23}|^2/\alpha_n$ denote the power split ratio for a branch. The condition for equal power distribution leads to the following recursion relation: 
\begin{equation} 
    \alpha_\mathrm{cross}\alpha_n \rho_n = \alpha_\mathrm{cross}\alpha_n (1-\rho_n) \alpha_\mathrm{cross}\alpha_{n-1} \rho_{n-1},
\end{equation} 
where the left-hand and right-hand sides represent the branched power at the $n$-th and the adjacent $(n-1)$-th MAC units, respectively, assuming unit power at the $n$-th unit's input port.

This expression simplifies under the assumption that the total loss remains constant for all branches ($\alpha = \alpha_\mathrm{cross}\alpha_n$): 
\begin{align} 
\rho_n = \alpha(1-\rho_n)\rho_{n-1}, \nonumber \\ 
\therefore \rho_n^{-1} = 1 + \alpha^{-1}\rho_{n-1}^{-1}. 
\end{align} 
Considering no through port at $n=1$, we set $\rho_1 = 1$. The general solution for the power split ratio is: \begin{equation} 
\rho_n(\alpha) = \begin{cases} 
n^{-1}, &  (\alpha = 1) \\
\alpha^{n-1}(1-\alpha)(1-\alpha^n)^{-1}, &  (\alpha < 1) 
\end{cases}, \label{eq:psr} 
\end{equation} 
which is a function of total loss $\alpha$, covering both ideal ($\alpha = 1$) and practical ($\alpha < 1$) cases.

In our design, we first optimize the waveguide crossing modules to achieve a maximum insertion loss of $\alpha_\mathrm{cross} = 99.38\%$ (0.027 dB). For the branch modules, we target an insertion loss of $\alpha_n^\mathrm{(tar)} = 98\%$ (0.088 dB), resulting in a total target loss of $\alpha = 97.40\%$ (0.115 dB). Using Eq. \eqref{eq:psr}, we obtain $\rho_1 = 1$, $\rho_2 = 0.4934$, $\rho_3 = 0.3246$, and $\rho_4 = 0.2402$, which slightly deviate from the ideal values of $1/n$. Finally, we set the target amplitudes for the adjoint optimization as $|S_{13}|^2 = \alpha_n^\mathrm{(tar)}(1-\rho_n)$ and $|S_{23}|^2 =\alpha_n^\mathrm{(tar)}\rho_n$.

\section{Details of the adjoint optimization}
The top panel of Fig. S2 outlines the inverse design process for the submodules. Using the adjoint state method, the gradient of the objective function (defined based on field components) with respect to the optical potential (i.e., the permittivity distribution) is computed through just two full-wave simulations: one forward and one backward. Additionally, practical fabrication constraints, such as lithographic resolution and binary material choices, are taken into account. Although the design parameter $q$ may vary abruptly across the design area, it is smoothed and binarized before being used in full-wave (FDTD) simulations to ensure that the permittivity distribution $\epsilon(x, y)$, derived as a function of $q$, meets the fabrication requirements. The gradients are calculated at each layer of the optimization process, from $q$ to the objective function $f$, and then multiplied in reverse direction to obtain the total gradient, $\nabla_q f$, based on the chain rule.

Specifically, $q$ is defined as a 2D pixelated design parameter with a resolution of $350 \times 350$ over a square region with a side length of $3.5~\mu\mathrm{m}$. This parameter is smoothed using a conical filter with a convolution kernel: 
\begin{equation} 
\mathrm{Kernel}(x,y) = \frac{3}{\pi r_\mathrm{cone}^2}\max\qty(0, 1-\frac{\sqrt{x^2+y^2}}{r_\mathrm{cone}}) 
\end{equation} 
with a radius of $r_\mathrm{cone} = 120$ nm. It is then binarized using a tanh-based function: 
\begin{equation} 
\epsilon_\mathrm{bi}(q; \beta) = \frac{\epsilon_\mathrm{Si}-\epsilon_\mathrm{SiO_2}}{2} \tanh\qty[\beta \qty(q-\frac{1}{2})] + \frac{\epsilon_\mathrm{Si}+\epsilon_\mathrm{SiO_2}}{2}, 
\end{equation} 
so that it falls within the permittivity range $[\epsilon_\mathrm{SiO_2}, \epsilon_\mathrm{Si}]$, corresponding to $\mathrm{SiO_2}$ and Si at the carrier frequency $f_0$. The binarization parameter $\beta$ is gradually increased from an initial value of 10 to 50 over the first 80 iterations, ultimately allowing for a fabricable layout in terms of the permittivity distribution. This design is applied to +-, L-, or T-shaped waveguide junctions, depending on the submodule type (Fig. 2b). Notably, the resulting footprint, on the order of microns, is much smaller than that of conventional integrated photonics, which typically requires waveguide coupling over several tens of microns, allowing for significant device size reduction.

The loss function for a single scattering parameter ($S$) relative to a target value $S' \in \mathbb{C}$ is given by: 
\begin{equation}
    \mathcal{L}(S(\mathbf{q}); S') = \begin{cases}
        |S-S'|^2, & (|S|\geq|S'|) \\ |S-S'|^2 + \eta (|S'|^2-|S|^2), & (|S|<|S'|)
    \end{cases},
\end{equation}
where $\eta \in [0,1]$ is a constant used to expedite the convergence by increasing the magnitude of $|S|$. When $\eta = 1$, for example, $\mathcal{L} = 2|S'|^2 - 2\mathrm{Re}(S^* S') = 2\mathrm{Re}[(S' - S)^* S']$, resulting in a sloped planar surface with a constant gradient within a circular region. On the other hand, setting $\eta = 0$ reduces the loss function to the standard squared $L^2$ distance. This approach is particularly useful when the target value $S'$ is large, such as $|S'| = 1$ for the crossing module, which is challenging to achieve due to physical limitations. In this study, we set $\alpha = 0.5$.

The total objective function for multi-port optimization (from input 3 to output $k = 1, 2, 3, 4$ for top, right, bottom, and left port, respectively) is defined as the weighted sum of the loss functions for each port: \begin{equation} 
    \mathcal{L}_\mathrm{tot}(\mathbf{S}(\mathbf{q}); \mathbf{S}') = \sum_k \gamma_k \mathcal{L}(S_{k,3}(\mathbf{q}); S_{k,3}'), 
\end{equation} 
where the weights $\gamma_k$ are chosen as follows: $\gamma = [1, 0.05, 0.05, 0.05]$ for the crossing, $[0.5, 0.5, 0.05, 0.05]$ for the beam splitter, $[0, 1, 0.05, 0]$ for the branch with $n=1$, and $[0.5, 0.5, 0.05, 0]$ for branches with $n \geq 2$.

\bibliography{references}

\newpage

\begin{figure}[h!]
    \centering
    \includegraphics[scale=1]{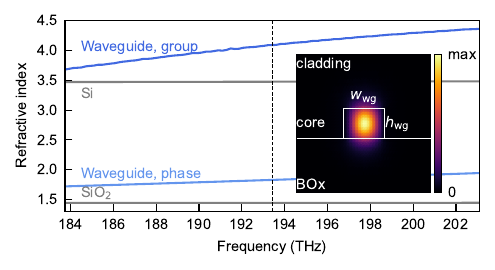}
    \caption{Refractive indices of crystalline silicon (c-Si) and silicon oxide (SiO$2$) \cite{material}, along with the corresponding effective group and phase indices for the TM${00}$ mode of the specified rectangular waveguide structure. The inset displays the mode profile for the TM$_{00}$ mode. The vertical dashed line marks the carrier frequency $f_0 = 193.4$ THz as the main text.}
    \label{fig:enter-label}
\end{figure}

\begin{figure}[h!]
    \centering
    \includegraphics[scale=1]{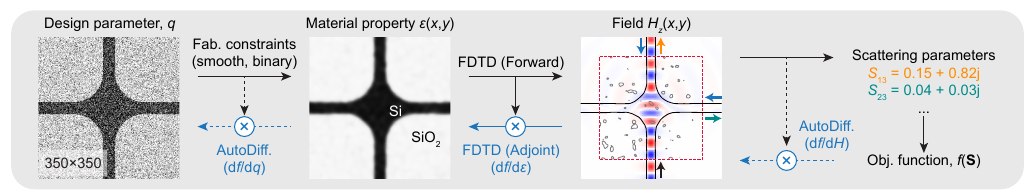}
    \includegraphics[scale=1]{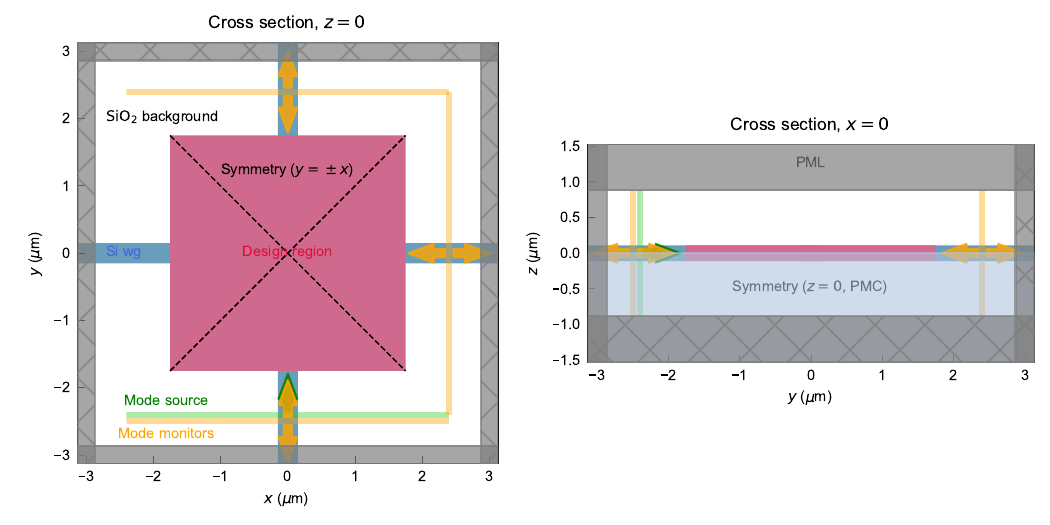}
    \caption{(Upper box) Inverse design procedure using adjoint-based optimization for optical structures.(Lower) Schematics of the FDTD simulation, showing cross sections at $z=0$ (left) and $x=0$ (right). The red box ($3 \times 3 \times 0.22~\mu\mathrm{m}^3$) indicates the design region, while the blue rods ($0.3 \times 0.22~\mu\mathrm{m}^2$ in cross-section) represent the Si waveguides. Green and yellow rectangular planes ($4.8 \times 1.77~\mu\mathrm{m}^2$) correspond to the mode source and monitors, respectively. Mirror symmetries are enforced along $y = \pm x$ in the design area, depending on the submodule type: $y = \pm x$ for the crossing and beam splitter, and $y = -x$ for the branch with $n=1$. Additionally, a $z=0$ PMC boundary condition is applied for all designs. The simulation space ($5.73 \times 5.73 \times 1.77~\mu\mathrm{m}^3$), with $\mathrm{SiO}_2$ as the background material, is surrounded by perfectly matched layers (PML).}
    \label{fig:enter-label}
\end{figure}

\begin{figure}[h!]
    \centering
    \includegraphics[scale=1]{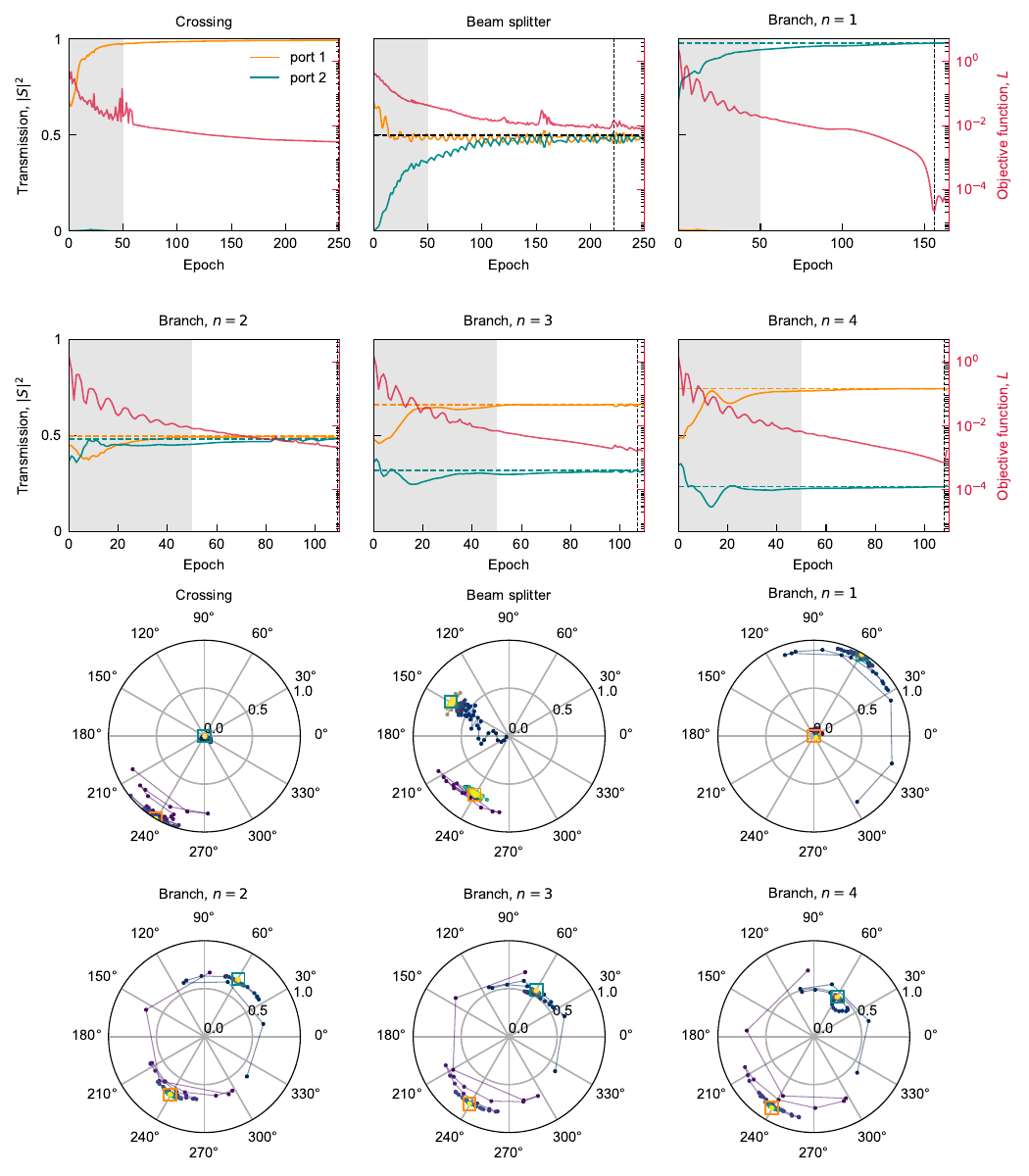}
    \caption{Learning curves of the adjoint optimization for the submodules: transmission coefficients ($|S|^2$, upper) and complex S-parameters ($S$, lower). The orange, teal, and red lines represent $|S_{13}|^2$ and $|S_{23}|^2$ for the mode monitors at the straight (through) and right-angle-bent waveguide, as well as the overall objective function, respectively. The orange and teal dashed lines in the upper plot and the square markers in the lower plot indicate the corresponding real and complex target values.}
    \label{fig:enter-label}
\end{figure}

\begin{figure}[h!]
    \centering
    \includegraphics[scale=1]{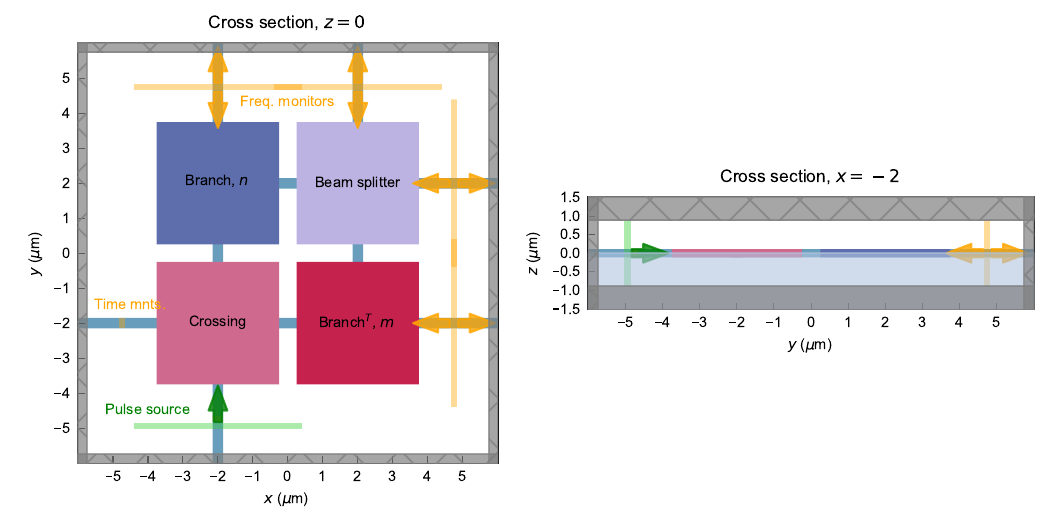}
    \caption{Schematics of the FDTD simulation for a single MAC unit, showing cross sections at $z=0$ (left) and $x=-2$ (right). Four colored boxes (red to purple), centered at $(x, y) = (\pm 2, \pm 2)$, represent the design area of each submodule. The blue waveguides are positioned at $x = \pm 2$ or $y = \pm 2$. Additional waveguide-sized field time monitors are placed at each input and output port to record the field dynamics. All other simulation conditions are identical to those in Fig. S2, unless otherwise specified.}
    \label{fig:enter-label}
\end{figure}

\begin{figure}[h!]
    \centering
    \includegraphics[scale=1]{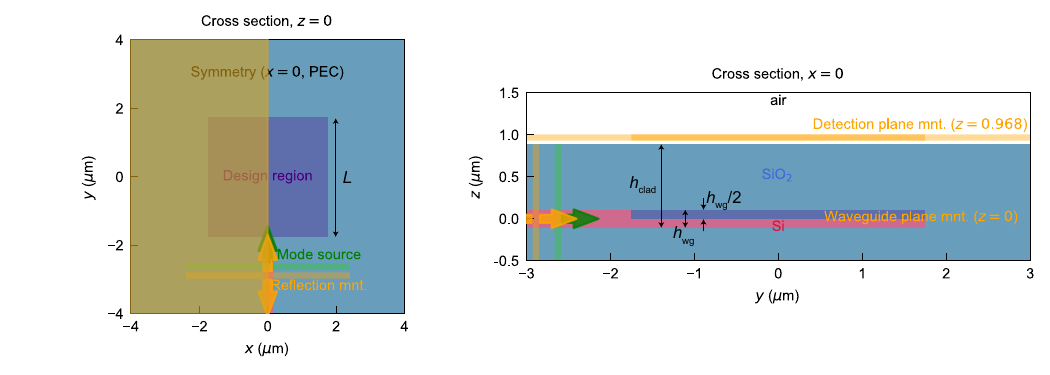}
    \caption{Design schematic of the small-area grating coupler: waveguide plane ($z=0$, left) and vertical cut plane ($x=0$, right). The square design area (purple) has a side length of $L = 3.5~\mu\mathrm{m}$ and a height of $h_\mathrm{wg}/2 = 0.11~\mu\mathrm{m}$. It is stacked on top of a silicon (Si) square base structure (pink) of the same size, which is connected to the Si waveguide. This structure is immersed in a silicon dioxide (SiO$2$) buried oxide region and cladding with a thickness of $h\mathrm{clad} = 0.78~\mu\mathrm{m}$. The detection plane is positioned at $z = 0.968~\mu\mathrm{m}$ in the air. A PEC boundary condition is applied at $x=0$ due to mirror symmetry. All dimensions are given in micrometers.}
    \label{fig:enter-label}
\end{figure}

\begin{figure}[h!]
    \centering
    \includegraphics[scale=1]{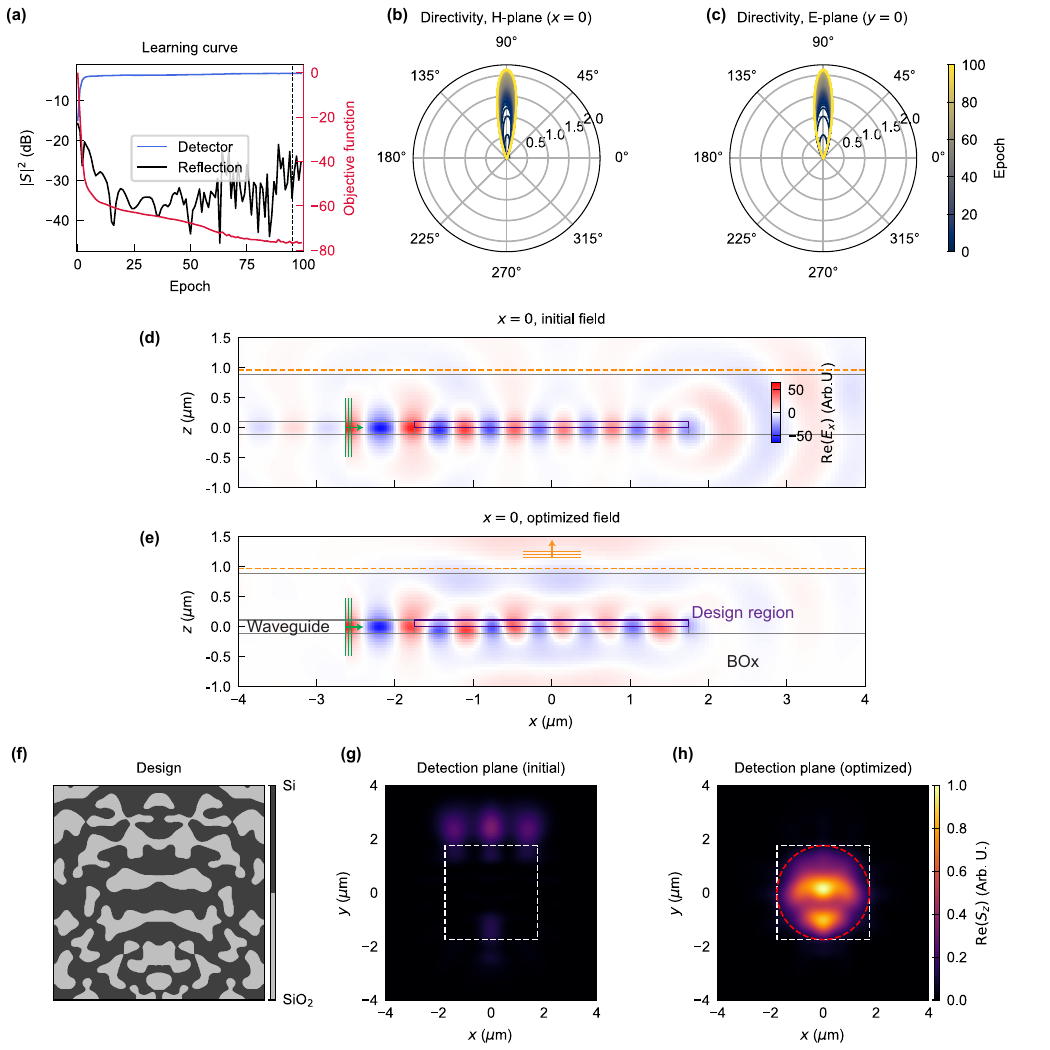}
    \caption{Optimization results for the small-area grating coupler. The objective function is defined as the coupling strength of the output field, $E_x(\mathbf{x}, y)$, measured at the detection plane (Fig. S5), with respect to the target Gaussian wave $\exp[-(x^2 + y^2)/w_0^2]$ with a beam waist of $w_0 = L/\sqrt{2}$. Using this objective function, the design region shown in Fig. S5 is optimized through the adjoint method. \textbf{a}-\textbf{c}, Learning curve (\textbf{a}) and the evolution of directivities in the H-plane ($x=0$, \textbf{b}) and E-plane ($y=0$, \textbf{c}). \textbf{d},\textbf{e}, Field ($E_x$) profiles in the $x=0$ plane for the initial and optimized structures, respectively. \textbf{f}, Optimized design layout. \textbf{g},\textbf{h} Out-of-plane power flux $\Re(S_z)$ measured at the detection plane (indicated by orange dashed lines in \textbf{d} and \textbf{e}) for the initial and optimized designs, respectively. The white dashed boxes denote the square design area, while the red dashed circle represents the beam waist ($L/2$) of the target Gaussian wave.}
    \label{fig:enter-label}
\end{figure}

\begin{figure}[h!]
    \centering
    \includegraphics[scale=1]{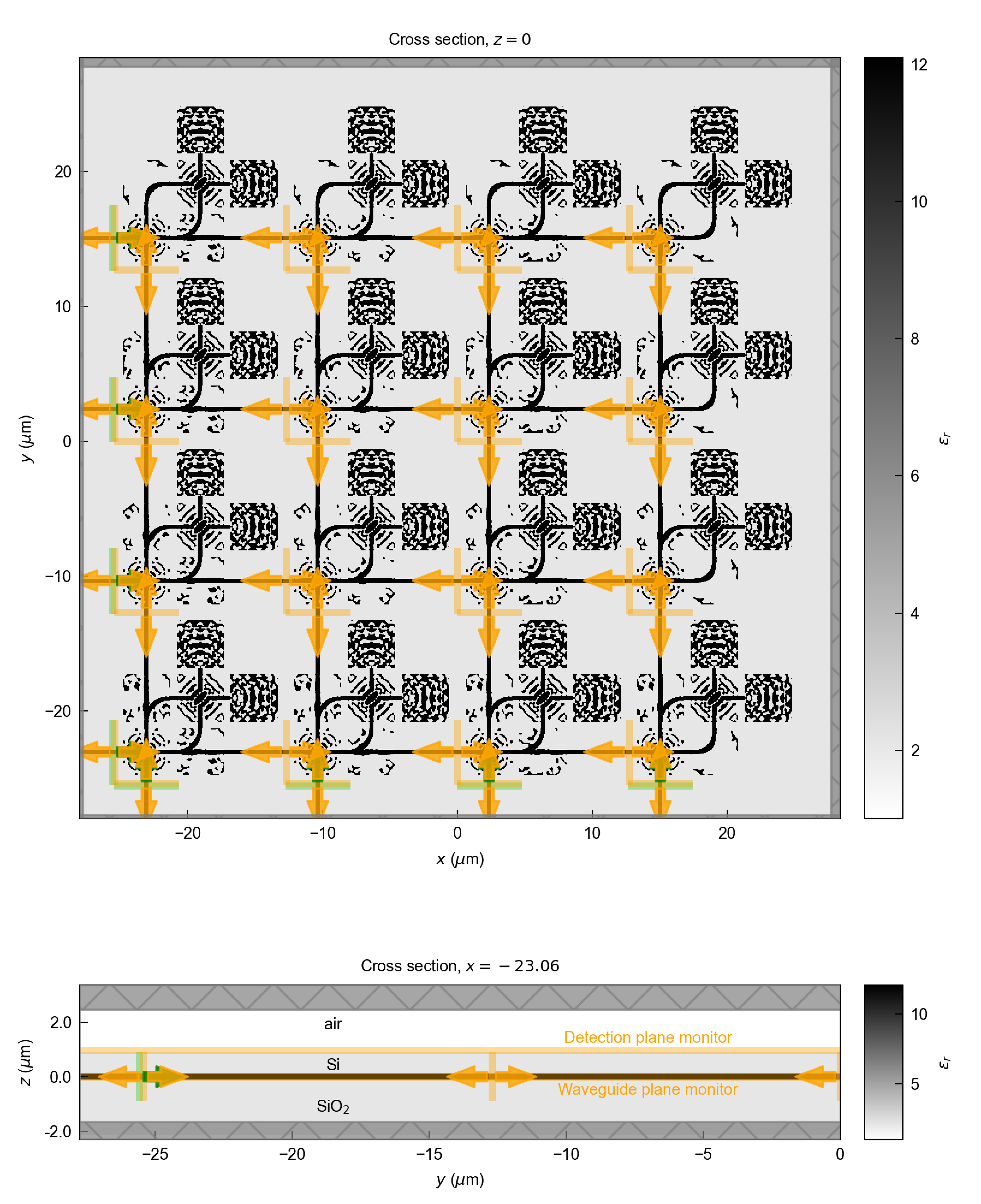}
    \caption{
    Simulation schematics for the $4 \times 4$ photonic systolic array: waveguide plane ($z=0$, left) and a vertical cross-section along one waveguide (right). The result for arbitrary input signals ($\vb A = [A_1, A_2, A_3, A_4]$ and $\vb B = [B_1, B_2, B_3, B_4]$) can be generated by superposing the result of single-source injections from any of the 8 sources. The lattice constant is set to 15 times the effective wavelength along the waveguide ($= 12.71~\mu\mathrm{m}$).
    }
    \label{fig:enter-label}
\end{figure}

\begin{figure}[h!]
    \centering
    \includegraphics[scale=1]{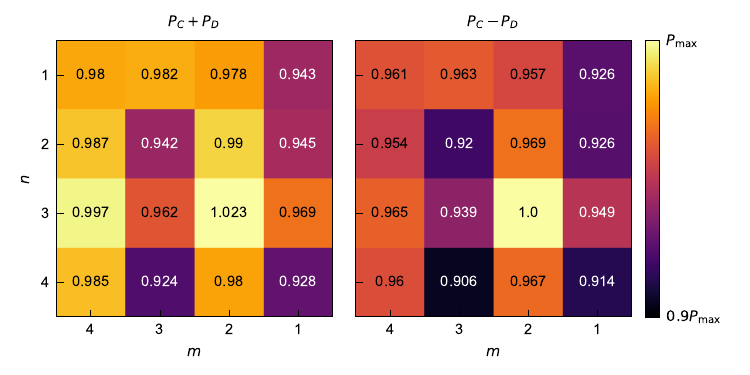}
    \caption{The sum (left) and difference (right) of the pulse powers detected at detectors $C$ and $D$ in the $(m,n)$-MAC unit when the inputs $A$ and $B$ are set to $A_m = 1$ and $B_n = 1$, with all other values being zero. All power measurements are given in units of $P_\mathrm{max} \equiv \max(P_C - P_D)$. The power difference matrix on the right is used for normalizing the outer product results, as discussed in the main text.}
    \label{fig:enter-label}
\end{figure}

\begin{figure}
    \centering
    \includegraphics[scale=1]{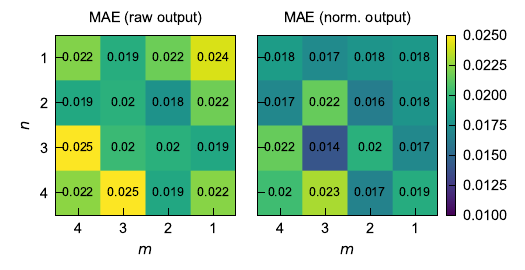}
    \includegraphics[scale=1]{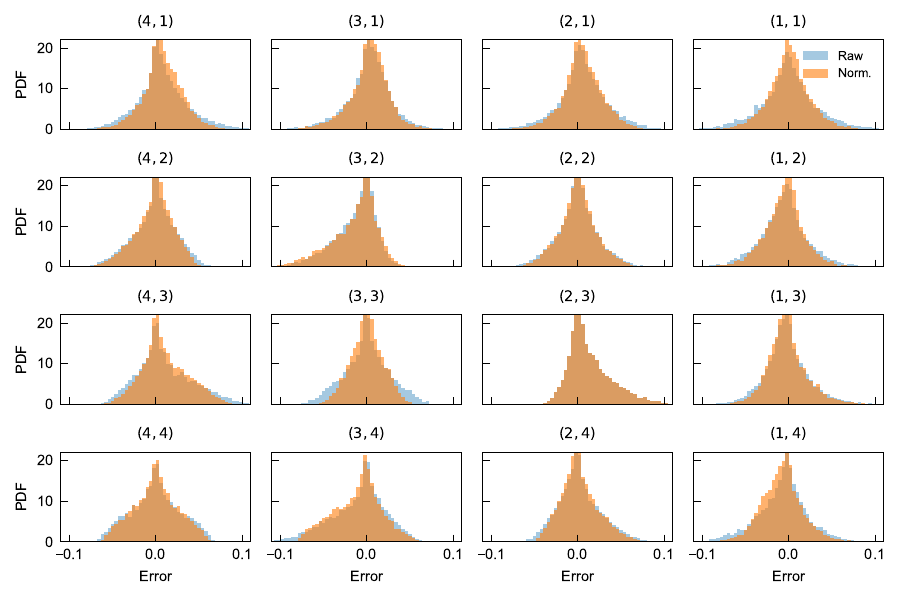}
    \caption{Monte Carlo simulation using a set of $10^4$ randomly sampled inputs to calculate the mean average error (MAE) for the output: the raw data MAE (upper left, total: 0.0212), the normalized output MAE (upper right, total: 0.0185), with respect to the ground truth. The lower 16 plots provide detailed error distributions for each individual MAC unit.}
    \label{fig:enter-label}
\end{figure}

\newpage

\appendix